\documentclass[]{aa}

\usepackage{txfonts}
\usepackage{graphicx}
\usepackage{natbib}
\usepackage{rotating}
\usepackage{url}
\usepackage{amssymb}
\bibpunct{(}{)}{;}{a}{}{,} 

\newcommand{\rah}{\textsuperscript{h}}
\newcommand{\ram}{\textsuperscript{m}}

\makeatletter
\def\artanh{\mathop{\operator@font artanh}\nolimits}
\makeatother
\begin{document}

\title{A Lensing Survey of X-ray Luminous Galaxy Clusters at Redshift
  $z \sim 0.2$}

\subtitle{II: CFH12k Weak Lensing Analysis and Global Correlations
  \thanks{Based on observations obtained at the Canada-France-Hawaii
    Telescope (CFHT) which is operated by the National Research
    Council of Canada, the Institut National des Sciences de l'Univers
    of the Centre National de la Recherche Scientifique of France, and
    the University of Hawaii.}  }

\titlerunning{Weak lensing survey of X-ray luminous galaxy clusters. II.}

\author{S. Bardeau\inst{1,2}
  \and
  G. Soucail\inst{1}
  \and
  J.-P. Kneib\inst{3,4}
  \and
  O. Czoske\inst{5,6}
  \and
  H. Ebeling\inst{7}
  \and
  P. Hudelot\inst{1,5}
  \and
  I. Smail\inst{8}
  \and
  G. P. Smith\inst{4,9}
}

\offprints{G. Soucail, soucail@ast.obs-mip.fr}

\institute{
  Laboratoire d'Astrophysique de Toulouse-Tarbes, CNRS-UMR 5572 and
  Universit\'e Paul Sabatier Toulouse III, 14 Avenue Belin,
  31400 Toulouse,  France
  \and
  Laboratoire d'Astrodynamique, d'Astrophysique et d'A\'eronomie de
  Bordeaux, CNRS-UMR 5804 and Universit\'e de Bordeaux I, 2 rue de
  l'Observatoire, BP 89, 33270 Floirac, France
  \and
  Laboratoire d'Astrophysique de Marseille, OAMP, CNRS-UMR 6110, 
  Traverse du Siphon, BP 8, 
  13376 Marseille Cedex 12, France
  \and
  Department of Astronomy, California Institute of Technology, Mail
  Code 105-24, Pasadena, CA 91125, USA
  \and
  Argelander-Institut f\"ur Astronomie, Universit\"at Bonn,
  Auf dem H\"ugel 71, 53121 Bonn, Germany
  \thanks{Founded by merging of the Sternwarte, Radioastronomisches
    Institut and Institut f\"ur Astrophysik und Extraterrestrische Forschung
    der Universit\"at Bonn}
  \and
  Kapteyn Astronomical Institute, P.O. Box 800, 9700 AV Groningen, The Netherlands
  \and
  Institute for Astronomy, University of Hawaii, 2680 Woodlawn Dr,
  Honolulu, HI 96822, USA
  \and
  Institute for Computational Cosmology, Department of Physics, 
  Durham University, South Road, Durham DH1~3LE, UK
  \and
  School of Physics and Astronomy, University of Birmingham,
  Edgbaston, Birmingham, B15~2TT, UK.
}

\date{Received 09/03/07; accepted 09/05/07}

\abstract
{}
{We present a wide-field multi-color survey of a homogeneous sample of
  eleven clusters of galaxies for which we measure total masses and mass
  distributions from weak lensing. }
{The eleven clusters in our sample are all X-ray luminous and span a
  narrow redshift range at $z=0.21 \pm 0.04$. The weak lensing
  analysis of the sample is based on ground-based wide-field
  imaging obtained with the CFH12k camera
  on CFHT. We use the methodology developed and applied previously on
  the massive cluster Abell\,1689. A Bayesian method, implemented in
  the \textsc{Im2shape} software, is used to fit the shape parameters
  of the faint background galaxies and to correct for PSF
  smearing. A multi-color selection of the background galaxies is
  applied to retrieve the weak lensing signal, resulting in a
  background density of sources of $\sim 10$ galaxies per square arc
  minute. With the present data, shear profiles are measured in all
  clusters out to at least 2~Mpc (more than 15\arcmin\ from the
  center) with high confidence. The radial shear profiles are fitted
  with different parametric mass profiles and the virial mass
  $M_{200}$ is estimated for each cluster and then compared to other
  physical properties.}
{Scaling relations between mass and optical luminosity indicate an
  increase of the $M/L$ ratio with luminosity ($M/L \propto L^{0.8}$)
  and a $L_{\mathrm{X}}-M_{200}$ relation scaling as
  $L_{\mathrm{X}} \propto M_{200}^{0.83 \pm 0.11}$ while the
  normalization of the $M_{200} \propto T_{\mathrm{X}}^{3/2}$ relation
  is close to the one expected from hydrodynamical simulations of
  cluster formation as well as previous X-ray analyses.  We
  suggest that the dispersion in the $M_{200}-T_{\mathrm{X}}$ and
  $M_{200}-L_{\mathrm{X}}$ relations reflects the different
  merging and dynamical histories for clusters of similar
  X-ray luminosities and intrinsic variations in their measured masses.
  Improved statistics of clusters over a wider mass range are required
  for a better control of the intrinsic scatter in scaling relations.}
{}


\maketitle

\section{Introduction}
\label{sec:introduction}
Clusters of galaxies are potentially powerful probes for cosmology.
They form the high-mass end of the mass function of collapsed halos, whose
development as a function of redshift is a basic test of the hierarchical
structure formation scenario and depends sensitively on a number of
cosmological parameters \citep{eke96,Voit2005}.  The distribution of
mass within clusters forms another test of the non-linear development of
structures \citep{navarro97}.  The main difficulty in the application of
these tests is to accurately measure total masses and the distribution
of mass in clusters from observational proxies which are only more or
less indirectly related to mass or the gravitational potential.

Several observational techniques are available to probe the mass
distribution in clusters, each of them based on different physical
assumptions and having its own strengths and weaknesses.  Observation
of the internal dynamics of clusters, based on the virial theorem and
using the cluster galaxies as test particles of the cluster potential,
is the ``historical'' approach, which provided early evidence for the
existence of ``missing'' (now ``dark'') matter \citep{zwicky37}. However,
clusters of galaxies are far from being simple relaxed systems and their
structural complexity makes analysis of the velocity field difficult,
as soon as the system shows substructure \citep{czoske02a}.

Due to its dependence on the square of the electron density, X-ray
emission from the hot intra-cluster gas (IGM) traces the deeper parts
of the cluster potential and can be used to infer the total cluster mass
under the assumption of hydrostatic equilibrium.  It is known, however,
that hydrostatic equilibrium alone provides an incomplete description of
the physics of the IGM. X-ray observations with \textit{XMM-Newton} and
\textit{Chandra} have revealed a wealth of complexity in the cluster X-ray
emission \citep{Markevitch-Vikhlinin2007} which sign-post deviations from
hydrodynamic equilibrium on various levels ranging from ``sloshing'' in
apparently relaxed clusters to gross deviations in, e.g., cluster mergers
\citep{finoguenov05,clowe06}.  \citet{Nagai2007} show from simulations
that mass estimates of relaxed clusters are biased low by 5 to 10\%
under the assumption that the gas is supported only by hydrostatic
pressure; for unrelaxed clusters the situation is much worse. Hence,
the usual X-ray observables like $L_{\mathrm{X}}$ or $T_{\mathrm{X}}$
do not provide the simple and expected robust mass estimators, although
the newly introduced $Y_{\mathrm{X}}$ parameter (i.e.~the product of
$T_{\mathrm{X}}$ and $M_{\mathrm{g},500}$, \citet{kravtsov06}) shows
some promise, with a low scatter when scaled with the mass.

The measurement of the Sunyaev-Zeldovich (SZ) effect is sensitive to the
integrated pressure of the intra-cluster gas and is potentially a robust
mass estimator \citep{grego01,bonamente06}.  The new SZ interferometers
like AMI \citep[AMI collaboration]{ami06} and millimeter bolometer
arrays (e.g.~LABOCA on APEX) are promising facilities for measuring the
SZ effect in clusters, although their limited spatial resolution will
limit the quality of cluster mass distribution.

Finally, weak gravitational lensing is most directly related to the
gravitational potential and hence the total mass distribution; in
particular, lensing does not rely on any assumptions concerning the
physical state of the system. The main systematic effect afflicting
lens mass measurements is that lensing measures projected masses,
hence the interpretation of lensing measurements in terms of physical
three-dimensional masses relies on certain assumptions concerning the
spatial distribution of the matter. Projection effects can range from
slight biases arising from triaxiality of the clusters or projection of
the general large-scale structure \citep{Metzler2001,
  King-Corless2007, Corless-King2007} to large errors in the mass
estimates in cases where there are unrecognized line-of-sight mergers of
clusters of comparable size \citep{czoske02a}.  Furthermore, there are
calibration issues in the mass measurements coming from 1)~the removal of
instrumental effects acting on the galaxy shapes, 2)~the calibration of
the weak lensing signal, 3)~the dilution of the signal by contamination
with faint cluster or foreground galaxies, and 4)~the uncertainties in
the source redshift distribution. The first two issues have largely been
addressed with simulated data in the STEP collaboration \citep[Shear
TEsting Program,][]{heymans06} aimed at comparing many different
weak lensing methodologies and identifying systematic effects in the
respective methods.  The last two issues can be addressed if additional
information is available such as multi-color photometry and photometric
redshifts or spectroscopic data.

From the above discussion it is clear that joint analyses of several
types of observations are necessary if one wants to fully understand
the structure of clusters of galaxies and the relation between different
observables and mass estimates \citep{dahle02,cypriano04}.  In practice,
such an in-depth analysis can only be conducted for comparatively small
samples of clusters, from which the relations and the scatter around
the relations have to be empirically calibrated.

The aim of our program is to study a homogeneous sample of galaxy clusters
using a variety of techniques to constrain their mass distribution.
\citet{smith05} presented the results from strong lensing mass modeling
of HST/WFPC2 observations of the central parts of the clusters and
a comparison to X-ray temperatures measured by \textit{Chandra}.
Analyses of \textit{XMM-Newton} observations of mostly the same clusters
are presented by \citet{zhang07}.  In this paper, we present the weak
lensing analysis of the cluster sample, using imaging data obtained
with the CFH12k wide-field camera \citep{cuillandre00} mounted at the
Canada-France-Hawaii Telescope (CFHT); eleven clusters were observed
in the $B$, $R$ and $I$ bands.  The weak lensing analysis of the full
sample of clusters follows the methodology presented previously by
\citet[][hereafter Paper~I]{bardeau05} who also applied it to the
cluster A\,1689.

This paper is organized as follows: Sect.~\ref{sec:sample} presents
the cluster sample and gives a summary of the data reduction and the
construction of catalogs that are used in the weak lensing and optical
analyses of the clusters.  In Sect.~\ref{sec:2d-mass-maps}, we convert
the galaxy shape measurements into two-dimensional mass maps which
allow a qualitative assessment of the morphology of the dark matter
distribution and comparison to the galaxy distribution in the clusters.
In Sect.~\ref{sec:wlmass}, we provide quantitative measurements of
the cluster masses using one-dimensional radial fits to the weak shear
signal.  Sect.~\ref{sec:optical} discusses the light distribution of
the clusters compared to the weak lensing mass in order to measure
the mass-to-light ratio.  We also compare the weak lensing masses to
the X-ray properties of the clusters (luminosity and temperature) and
normalize the $M$--$L_{\mathrm{X}}$ and $M$--$T_{\mathrm{X}}$ relations.
We summarize our conclusions in Sect.~\ref{sec:conclusions} and discuss
prospects for future weak lensing cluster surveys.  Finally we give some
details on the individual properties of the clusters in the appendix.

Throughout the paper we use $H_0 = 70\,\mathrm{km\,s^{-1}\,Mpc^{-1}}$,
$\Omega_\mathrm{M}=0.3$, $\Omega_\Lambda=0.7$. At a redshift of
$z=0.2$, $1\arcsec$ corresponds to $3.3\,\mathrm{kpc}$ and $1\arcmin$
to $200\,\mathrm{kpc}$. Magnitudes are given in the Vega system.

\section{Observations and data reduction}
\label{sec:sample}

\subsection{Description of the cluster sample}
\label{ssec:descr-clust-sample}
\begin{figure}
  \centering
  \resizebox{\hsize}{!}{\includegraphics{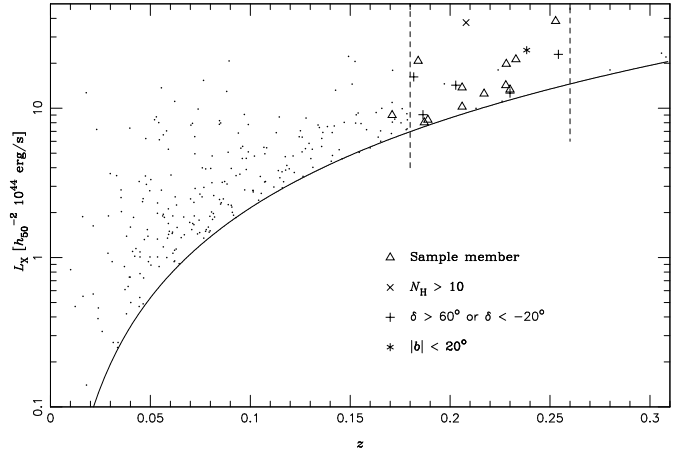}}
  \caption{$L_\mathrm{X}$--$z$ diagram for clusters from the XBACs
    catalog (dots).  Triangles mark the 12 members of our sample,
    although one cluster (A\,773) could not be observed and is
    not studied in this paper. The other
    symbols mark clusters which failed some of our secondary selection
    criteria. The solid line corresponds to an X-ray flux of $5 \times
    10^{-12}\,\mathrm{erg\,s^{-1}\,cm^{-2}}$, the flux limit of XBACs.
    The point outside the redshift limits (dashed lines) corresponds
    to the cluster A2218. Luminosities given here are the original
    XBACs luminosities from \textsc{Rosat} All-Sky Survey data
    computed for $H_0 = 50\,\mathrm{km\,s^{-1}\,Mpc^{-1}}$,
    $\Omega_{\mathrm{M}}=1$, $\Omega_{\Lambda} = 0$.}
  \label{fig:xbacs}
\end{figure}

The clusters analyzed in this paper were selected from the XBACs catalog
of \citet{ebeling96}, a flux-limited compilation of Abell clusters
detected in the \textsc{Rosat} All-Sky Survey data.  While the sample
is thus based on the optically selected Abell catalog \citep{abell89},
which is known to be incomplete at high redshift and low mass, comparison
with the X-ray selected BCS \citep{ebeling98,ebeling00} shows that
more than 74\% of all BCS clusters are indeed Abell clusters. Since,
for the very X-ray luminous systems considered here, this fraction rises
to almost 90\%, the optical preselection should not bias our sample. We
select systems within the relatively narrow redshift range of $0.17 <
z < 0.26$ in order to obtain an approximately luminosity-limited sample
and to ensure that all clusters lie at about the same distance from the
background population. Since X-ray luminosity is broadly correlated with
cluster mass \citep{reiprich02}, our sample is approximately mass limited,
too. The redshift range quoted above was chosen to maximize the lensing
efficiency for a background galaxy population at $\langle z\rangle \sim
1.0$ \citep{natarajan97}, thus defining the requirements in terms of
limiting magnitudes.  We applied further limits in declination ($ -20\degr
< \delta < 60\degr$ for accessibility with CFHT from Mauna Kea), Galactic
latitude ($|b| > 20\degr$ to minimize contamination by stars) and hydrogen
column density ($N_\mathrm{H} < 10 \times 10^{20}\,\mathrm{cm^{-2}}$)
and from the remaining clusters randomly selected a sample of twelve
clusters within a redshift dispersion $\sigma_z/z = 12\%$.  Of these,
eleven clusters were observed with the CFH12k wide-field camera (A773
could not be observed within the allocated nights).  The same cluster
sample was also imaged with the Hubble Space Telescope (HST) in order to
measure the mass distribution in the cluster cores using strong-lensing
techniques. Eight systems were observed by us with the WFPC2 camera
(Program ID: 8249, PI Kneib), and data from observations of another two
were retrieved from the HST archive \citep[see][]{smith05}.  Note that
sample completeness is not of critical importance for this project which
rather aims at compiling a homogeneous data set for a representative
comparison among mass measurement techniques.  Fig.~\ref{fig:xbacs} shows
how our sample covers the high-luminosity region in the XBACs catalog.

Table~\ref{tab:sample} provides a summary of the cluster properties
including global X-ray characteristics that will be used in our
analysis.

\begin{table*}
  \caption{Physical properties of the eleven clusters of the sample
    studied in this paper. The coordinates are those of the central
    galaxy as measured on the CFH12k images. The X-ray luminosities
    are taken from XMM-Newton measurements in \citet{zhang07} and are
    X-ray bolometric luminosities excluding the $<0.1 r_{500}$ region. 
    The X-ray temperatures listed in column~6 were measured
    from Chandra data by \citet{smith05}, excluding the very central
    regions where cool cores might be present in some of the clusters,
    except for A1689$^{*}$ \citep[XMM data]{andersson04} and
    A2390$^{**}$ \citep{allen01a}. Column 7 lists the X-ray
    temperature measured with XMM-Newton from \citet{zhang07}, except
    for A\,2219$^{***}$ for which we list the temperature obtained with ASCA
    \citep{ota04} as no XMM data are available. }
  \label{tab:sample} 
  \centering{
    \begin{tabular}{lcccccc}
      \hline\hline \noalign{\smallskip} Cluster &   RA  & Dec  & $z$ &
      $L_\mathrm{X}$ (XMM) & $T_\mathrm{X}$ (Chandra) & $T_\mathrm{X}$ (XMM) \\
      & (J2000)    & (J2000) &        &
      $10^{44}\, h_{70}^{-2} \ \mathrm{erg\,s}^{-1}$ & keV & keV \\
      \noalign{\smallskip} \hline \noalign{\smallskip}
{A\,68}   
      & 00\rah37\ram06\fs9 &  $+$09\degr09\arcmin24\arcsec
      & 0.255 & $10.1 \pm 0.9$ & $9.5\ ^{+1.5}_{-1.0}$ & $7.7 \pm 0.3$\\
      \noalign{\smallskip}
{A\,209}   
      & 01\rah31\ram52\fs6 &  $-$13\degr36\arcmin40\arcsec
      & 0.206 & $13.2 \pm 1.1$ & $8.7\ ^{+0.6}_{-0.5}$ & $7.1 \pm 0.3$\\
      \noalign{\smallskip}
{A\,267}   
      & 01\rah52\ram42\fs0 &  $+$01\degr00\arcmin26\arcsec
      & 0.230 & $6.6 \pm 0.7$ & $6.0\ ^{+0.7}_{-0.5}$ & $6.5 \pm 0.4$\\
      \noalign{\smallskip}
{A\,383}   
      & 02\rah48\ram03\fs4 &  $-$03\degr31\arcmin45\arcsec
      & 0.187 & $4.6 \pm 0.5$ & $5.2\ ^{+0.2}_{-0.2}$ & $5.3 \pm 0.2$\\
      \noalign{\smallskip}
{A\,963}   
      & 10\rah17\ram03\fs6 &  $+$39\degr02\arcmin50\arcsec
      & 0.206 & $10.2 \pm 0.9$ & $7.2\ ^{+0.3}_{-0.3}$ & $6.3 \pm 0.2$\\
      \noalign{\smallskip}
{A\,1689}  
      & 13\rah11\ram30\fs1 &  $-$01\degr20\arcmin28\arcsec
      & 0.184 & $21.4 \pm 1.0$ & $9.0\ ^{+0.13}_{-0.12} \
      ^{*}$ & $8.4 \pm 0.2$\\
      \noalign{\smallskip}
{A\,1763}  
      & 13\rah35\ram20\fs1 &  $+$41\degr00\arcmin04\arcsec
      & 0.228 & $15.9 \pm 1.4$ & $7.7\ ^{+0.4}_{-0.4}$ & $6.3 \pm 0.3$\\
      \noalign{\smallskip}
{A\,1835}  
      & 14\rah01\ram02\fs1 &  $+$02\degr52\arcmin42\arcsec
      & 0.253 & $30.0 \pm 1.4$ & $9.3\ ^{+0.6}_{-0.4}$ & $8.0 \pm 0.3$\\
      \noalign{\smallskip}
{A\,2218}  
      & 16\rah35\ram51\fs5 &  $+$66\degr12\arcmin15\arcsec
      & 0.171 & $12.2 \pm 0.9$ & $6.8\ ^{+0.5}_{-0.5}$ & $7.4 \pm 0.3$\\
      \noalign{\smallskip}
{A\,2219}  
      & 16\rah40\ram19\fs9 &  $+$46\degr42\arcmin41\arcsec
      & 0.228 & --- & $13.8\ ^{+0.8}_{-0.7}$ & $9.2 \pm 0.4 \ ^{***}$\\
      \noalign{\smallskip}
{A\,2390} 
      & 21\rah53\ram36\fs9 &  $+$17\degr41\arcmin43\arcsec
      & 0.233 & $28.9 \pm 2.2$ & $11.5\ ^{+1.6}_{-1.3} \
      ^{**}$ & $10.6 \pm 0.6$\\
      \noalign{\smallskip}\hline
    \end{tabular} 
  }
\end{table*}

\subsection{Observations}
\label{ssec:observations}
Imaging data were obtained at the Canada-France-Hawaii Telescope
during three observing runs in February 1999, November 1999
and May/June 2000 using the CFH12k camera, a mosaic of twelve
$2\mathrm{k}\,\times\,4\mathrm{k}$ CCDs \citep{cuillandre00}. With
a pixel scale of $0\farcs206$ this camera covers a field of view of
$42 \times 28\,\mathrm{arcmin^2}$, i.e.\ about 1/3 of a square degree.
Observations were conducted in three filters, $B$, $R$ and $I$. A1835
was observed in $V$ instead of $B$ in February 1999 because the latter
filter was not yet available at the time. Further observational details
are given in Table~\ref{tab:observations}.

We determine the limiting magnitude of each final image by computing the
magnitude of a point source detected at $5\sigma$ in an aperture with
diameter 1.45 times the full width at half maximum (FWHM) of the seeing
disk. As shown by the CFH12k exposure time calculator (DIET\footnote{
  \url{http://www.cfht.hawaii.edu/Instruments/Imaging/CFH12K/DIET/CFH12K-DIET.html}})
this is the optimal aperture within which 96\%\ of the flux is integrated
while the noise level remains sufficiently low. A point-like object
detected at this significance level will have a magnitude error
of about 20\%. The resulting limiting magnitudes are included in
Table~\ref{tab:observations}. The \emph{measured} magnitude limits
corresponding to 50\% completeness are found to be about 1.8 magnitudes
brighter, partly due to the small apertures used in DIET.

During all three runs, the observing conditions were very good in
terms of seeing (a prime requirement for weak-lensing studies) with
the FWHM of stars consistently below $1\arcsec$ in the $R$ band. The
first and second observing runs were photometric; the third run was
affected by cirrus (see Sect.~\ref{ssec:data-reduction}).

\begin{table*}
  \caption{Summary of the CFHT observations of our cluster sample.
    Exposure times are given in seconds. The seeing is measured as the
    FWHM of stars in the final stacked image for each cluster.  The
    limiting magnitudes were computed following the DIET recipes (see
    text for details) for a point source detected at $5\sigma$ in an
    area of diameter 1.45 times the FWHM of the image.  At
    this limit magnitudes are measured with an intrinsic error
    $\Delta m \sim 0.2$. They do not include the correction for
    galactic absorption which is given for   
    each cluster in the 3 filters.} 
  \centering
  \begin{tabular}{l|cccc|cccc|cccc}
    \hline\hline
    \raisebox{0ex}[2.5ex][0ex] & \multicolumn{4}{c|}{B} 
    & \multicolumn{4}{c|}{R} & \multicolumn{4}{c}{I} \\
    Cluster & Integr. & Seeing & Limiting & $A_B$ & 
              Integr. & Seeing & Limiting & $A_R$ &
              Integr. & Seeing & Limiting & $A_I$ \\
    \raisebox{0ex}[2ex][1ex] & time &  & mag. & & 
                               time &  & mag. & &
                               time &  & mag. \\ 
    \hline
    A\,68   \raisebox{0ex}[2.5ex][0ex]{} 
    & 8100     & 1\farcs1 & 26.4 & 0.40  & 7200 & 0\farcs7 & 26.3 & 0.25 
    & 3600  & 0\farcs6 & 25.3 & 0.18 \\
    A\,209  
    & 7200     & 1\farcs0 & 26.8 & 0.08  & 6600 & 0\farcs7 & 26.3 & 0.05
    & 3600  & 0\farcs7 & 25.1 & 0.04 \\
    A\,267  
    & 3000     & 1\farcs0 & 26.0 & 0.11  & 4800 & 0\farcs7 & 25.9 & 0.07
    & 900   & 0\farcs7 & 24.5 & 0.05 \\
    A\,383  
    & 7200     & 0\farcs9 & 27.0 & 0.14  & 6000 & 0\farcs9 & 26.1 & 0.09
    & 3600  & 0\farcs7 & 25.1 & 0.06 \\
    A\,963  
    & 7200     & 0\farcs9 & 27.0 & 0.06  & 4800 & 0\farcs8 & 26.1 & 0.04
    & 10500 & 1\farcs1 & 24.6 & 0.03 \\ 
    A\,1689 
    & 3600     & 0\farcs9 & 26.7 & 0.12  & 3000 & 0\farcs8 & 26.1 & 0.07
    & 3000  & 0\farcs9 & 24.8 & 0.05 \\  
    A\,1763 
    & 3600     & 1\farcs0 & 26.7 & 0.04  & 6000 & 0\farcs9 & 26.2 & 0.02
    & 3000  & 0\farcs8 & 25.0 & 0.02 \\ 
    A\,1835 
    & 3750 $^{*}$ & 0\farcs8 & 26.4$^{*}$ & 0.10$^{*}$ & 5400 & 0\farcs7 
& 26.5 & 0.08
    & 3750  & 0\farcs8 & 25.5 & 0.06 \\ 
    A\,2218 
    & 3378     & 1\farcs1 & 26.3 & 0.11  & 6900 & 1\farcs0 & 26.2 & 0.07
    & 3000  & 0\farcs8 & 24.7 & 0.05 \\ 
    A\,2219 
    & 5400     & 1\farcs0 & 26.8 & 0.11  & 6300 & 0\farcs8 & 26.4 & 0.07
    & 3000  & 0\farcs8 & 25.1 & 0.05 \\ 
    A\,2390  \raisebox{0ex}[2ex][1ex]
    & 2700     & 1\farcs1 & 26.2 & 0.48  & 5700 & 0\farcs7 & 26.3 & 0.30
    & 3600  & 0\farcs9 & 25.1 & 0.22 \\ 
    \hline
  \end{tabular} 

  \smallskip
  $^{*}$ V-band data for A\,1835. 
  \label{tab:observations}
\end{table*}

\subsection{Data reduction}
\label{ssec:data-reduction}
Full details of the data reduction are presented in \citet{czoske02b}.
In Paper~I we give a summary of the reduction process with special
emphasis on the astrometric alignment of the exposures, which is of
particular importance for the weak-lensing analysis of the images.  We
here provide additional information on the absolute photometric
calibration of the images which will be relevant for the comparison of
the total galaxy cluster luminosity to the total mass distribution
presented in Sect.~\ref{sec:optical}.

Before combining the exposures taken through the same filter for a
given cluster we apply empirically determined photometric scaling
factors to the exposures, which take into account differences in air
mass and atmospheric transparency. The factors are determined by
comparing the instrumental fluxes of several thousand stars in each
image to the corresponding fluxes in a reference exposure (usually the
exposure taken at the lowest air mass). The median of the distribution
of flux ratios is selected and used as the global scale factor which
brings the exposure to the same level as the reference exposure.
Remaining variations in the photometric zero point between chips
(which have not been completely corrected by the twilight flat fields)
are determined in a simpler manner using the sky levels in adjacent
parts of neighboring chips. The intrinsic accuracy of the photometric
calibration within the images is estimated to be on the order of
$0.01\,\mathrm{mag}$.

Plotting the relative scaling factors for the individual exposures against
air mass allows the determination of the atmospheric extinction if the
series of exposures has been taken over a sufficient range in air mass;
this is the case for several of our fields. We confirm that for the
first and second of our observing runs the conditions were photometric
throughout; the extinction coefficients determined in this way are in
good agreement with the values listed on the CFH12k web site\footnote{
  \url{http://www.cfht.hawaii.edu/Instruments/Imaging/CFH12K/Summary/CFH12K-Optics.html}}.
The photometric zero points are determined from observations of
photometric standard fields taken throughout the nights. We use
the catalogs by \citet{landolt92} supplemented more recently
by \citet{stetson00}. Finally, the Galactic extinction is
computed for each cluster from the maps of \citet{schlegel98},
accessed through the NASA/IPAC Extragalactic Database
(NED)\footnote{\url{http://nedwww.ipac.caltech.edu/}}. Extinction
corrections in R are typically on the order of 0.05\,mag with the
exception of Abell\,68 and Abell\,2390 for which we find values of up
to 0.30\,mag.

The third observing run was affected by intermittent thin cirrus and the
conditions were not photometric.  Calibrating the images taken during
this run with standard observations is therefore likely to result in
systematic errors in the photometric zero points. For the purposes of the
present paper we adjust the zero points \emph{a posteriori} by selecting
stars from our fields and comparing their color distribution to external
multi-color photometric sequences \citep{saha05}.  In the $R-I$ vs.~$B-R$
color plot the stellar sequence displays a characteristic knee from which
photometric color corrections in $B-R$ and $R-I$ can be determined. A
correction of 0.1 to 0.4 magnitudes is required for the clusters A1689,
A1763, A2218, A2219 and A2390 to obtain a correct value for the $R-I$
color of elliptical galaxies. The correction terms are arbitrarily
applied relative to the R-band which is taken as reference. Note that
except for the color selection of cluster ellipticals or of ``red''
galaxies (see below for their definition), only the R-band luminosities,
which are only mildly affected by the non-photometric conditions, are
used in the rest of this paper. Only the clusters A\,1763 and A\,2219
suffered from significant absorption during the R-band observations,
with an unknown absorption factor not higher than 0.3\,mag.

\subsection{Catalogs}
\label{ssec:catalogues}
From the corrected images, photometric catalogs are created for each
cluster using \textsc{SExtractor} \citep{bertin96}.  Different
catalogs are produced for the various measurements needed for the
analysis.  The details of the general methodology are given in Paper~I
and are summarized below.  Object colors are computed using aperture
magnitudes in circles of 3\arcsec\ diameter, and stellar objects are
identified using the tight relation in the magnitude vs.~peak
surface-brightness diagram.

\begin{figure}
  \centering
  \resizebox{\hsize}{!}{\includegraphics{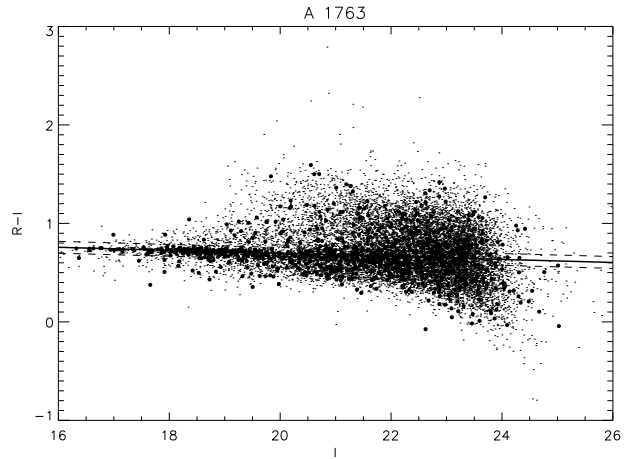}}
  \caption{Color-magnitude diagram for the cluster Abell~1763 showing
    the characteristic red sequence of elliptical cluster
    galaxies (width $\Delta m = \pm 0.06$). 
    All galaxies redder than the sequence are included in
    the ``red'' galaxy catalog. The large dots correspond to galaxies
    within a radius of 250\arcsec\ while the small dots come from the
    full catalog.}
  \label{fig:colmag}
\end{figure}

Color-magnitude diagrams are constructed from the photometric catalog
(Fig.~\ref{fig:colmag}), and the cluster red sequence is manually
identified and approximated by a line with negative slope; all galaxies
in a certain range around the line ($\Delta m = \pm 0.08$ typically) are
stored in a catalog which we will refer to as the ``elliptical galaxy''
catalog. Since, for simplicity's sake, we do not introduce any cut in
magnitude, this catalog is affected by field contamination, especially
at faint magnitudes.

For each cluster the photometric catalog is split into two sub-catalogs
containing ``bright'' and ``faint'' galaxies, respectively. The
magnitude cuts used to select these catalogs are scaled to the
value of $m^{\star}$, determined separately for each cluster and
each filter (see Paper~I for details and Table~\ref{tab:photom}).
The ``faint galaxy'' catalog is dominated by background sources
and is used for the weak-lensing analysis. The galaxy density
ranges from $22~\mathrm{galaxies}/\mathrm{arcmin^2}$ in $R$ to
$14~\mathrm{galaxies}/\mathrm{arcmin^2}$ in $B$ and $I$.  These numbers
are averaged densities over the cluster sample and differ slightly from
cluster to cluster due to cosmic variance and/or differences in the
photometric depth of the catalogs (Table~\ref{tab:photom}).

\begin{table}
  \caption{Photometric properties of the sub-catalogs used in the
    weak-lensing analysis. Only R-band data are summarized. The
    magnitude range used for the faint galaxy selection is defined to
    be: [$m^\star_R +3.5$ to $m_\mathrm{comp}+0.5$], $m_\mathrm{comp}$
    being the completeness magnitude of the catalog.}
  \label{tab:photom}
  \centering
  \begin{tabular}{lccc}
    \hline\hline\noalign{\smallskip}
    Cluster & $m^\star_R$ &  Faint galaxies & Mean number density \\
    & &  magnitude range & (gal / arcmin$^2$) \\
    \noalign{\smallskip}\hline\noalign{\smallskip}
    A\,68   & 19.04 & [22.5--24.9] & 22 \\
    A\,209  & 18.41 & [21.9--24.9] & 24 \\
    A\,267  & 18.71 & [22.2--24.5] & 17 \\
    A\,383  & 18.12 & [21.6--24.9] & 22 \\
    A\,963  & 18.41 & [21.9--24.8] & 23 \\
    A\,1689 & 18.08 & [21.6--24.8] & 22 \\
    A\,1763 & 18.67 & [22.2--25.0] & 24 \\
    A\,1835 & 19.02 & [22.5--25.4] & 25 \\
    A\,2218 & 17.91 & [21.4--25.0] & 21 \\
    A\,2219 & 18.67 & [22.2--25.3] & 25 \\
    A\,2390 & 18.75 & [22.3--24.6] & 18 \\
    \noalign{\smallskip}\hline
  \end{tabular}
\end{table}

\begin{figure}
  \centering
  \resizebox{\hsize}{!}{\includegraphics{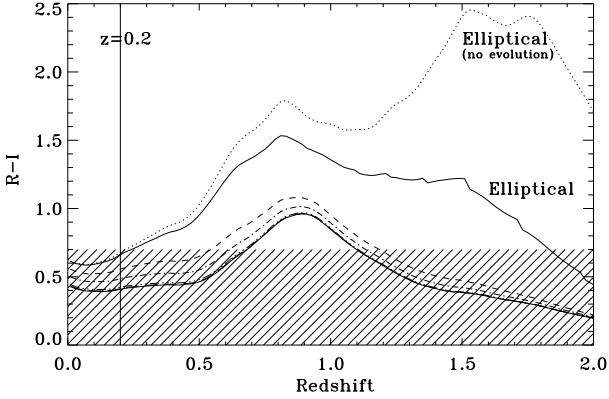}}
  \caption{Observed color index $R-I$ versus redshift for different
    spectral types, ranging from elliptical (with and without
    evolution, marked by full and dotted lines, respectively) to the
    bluest star-forming galaxies.  We select ``red galaxies'' with
    $R-I > 0.7$ (outside the hatched region): The resulting catalog is
    next to free from foreground galaxies and cluster members, and is
    limited to $z \lesssim 1.8$.  Synthetic spectral models are from
    \citet{bruzual03}.}
  \label{fig:RIz}
\end{figure}

\begin{figure}
  \centering
  \resizebox{\hsize}{!}{\includegraphics{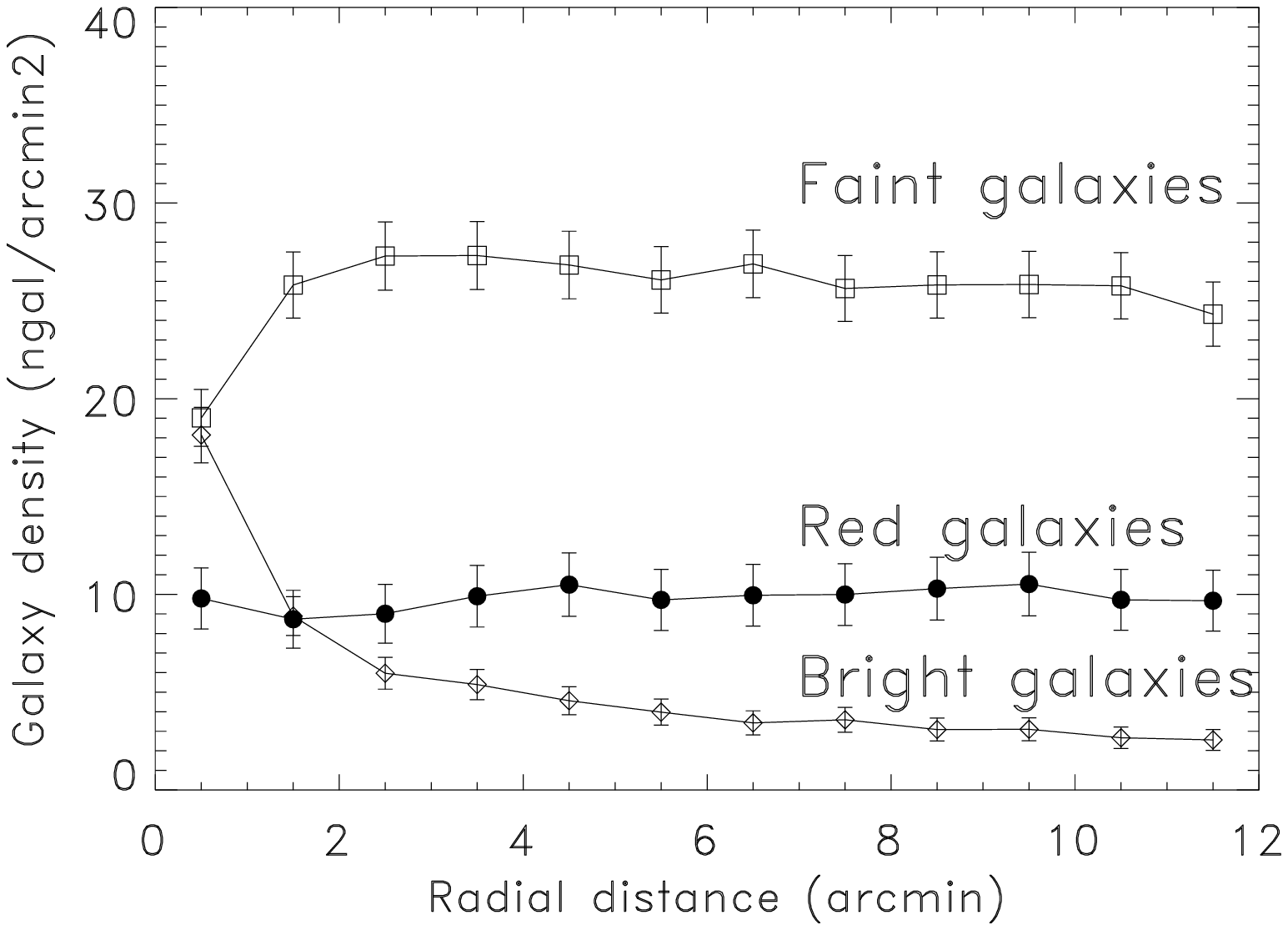}}
  \caption{Mean radial profile of the ``red'' galaxies, averaged over
    the four ``relaxed'' clusters (A\,383, A\,963, A\,1689, and
    A\,1835).  Also plotted is the mean radial profile for the
    ``bright'' galaxies, representing the cluster members
    complementing the ``faint'' galaxy sample. As expected the ``red''
    galaxy profile is flat and therefore mostly free from cluster
    contamination. The central dip in the faint galaxies profile may
    be partly due to the magnification bias acting on the background
    sources and also to obscuration effects caused by contaminating
    bright galaxies (not corrected).}
  \label{fig:radprof}
\end{figure}

To further select faint background galaxies as well as to minimize cluster
contamination, we add a simple color criterion to isolate galaxies above
the cluster red sequence.  As shown in Fig.~\ref{fig:RIz}, galaxies
redder than $R-I \ga 0.7$ should essentially be background galaxies
ranging from the cluster redshift up to $z \simeq 1.5-1.8$. This
approach has already been successfully developed and used by
several groups \citep{kneib03,broadhurst05b} to eliminate or at
least severely reduce the contamination of background galaxy samples
by cluster galaxies. This extra criterion reduces the galaxy number
densities to typically $8-10\,\mathrm{galaxies}/\mathrm{arcmin^2}$ (a
reduction by a factor two compared to the ``faint galaxy'' catalogs,
see Fig.~\ref{fig:radprof}). However, because the ``red galaxy''
catalog can be assumed to contain mostly background sources, we use
this catalog for the remainder of our weak-lensing analysis, except in
Sect.~\ref{sec:2d-mass-maps} where for practical reasons the ``faint
galaxy'' catalogs are used as input.

\subsection{Measurement of galaxy shape parameters}
\label{ssec:shape_parameters}
The shape parameters of faint galaxies are measured in all images,
allowing a qualitative comparison between the different bands.  However,
in the end we only retain the measurements in the $R$ band because
the data quality in this band in terms of seeing and source density is
superior to that of the other bands. In order to measure the galaxy shapes
for the weak-lensing analysis, we correct them for PSF smearing using
the \textsc{Im2shape} software developed by \citet{bridle01}. Details
of the implementation are given by \citet{heymans06} as well as a
comparison with more commonly used techniques such as the KSB~scheme
\citep{kaiser95}. Briefly, \textsc{Im2shape} fits each object image with
two concentric elliptical Gaussians convolved with a local estimate
of the PSF, taking into account the background and noise levels.
The local PSF is determined as the average shape of the three stars
closest to each object.  Shape parameters and ellipticities with their
error estimates are measured from the deconvolved object images using
a Monte Carlo Markov Chain (MCMC) optimization technique.

\subsection{Total optical luminosities of the clusters}
\label{sec:total-luminosity}
There are several ways to estimate the total optical luminosities of
the clusters from our wide-field images. In all cases, absolute
luminosities include a $k$-correction obtained at the cluster redshift
for an elliptical-type spectral energy distribution \citep{bruzual03}.

First, the total luminosity is integrated in circular annuli from the
``bright galaxy'' catalog and corrected for an average background
luminosity computed in the external areas of the images.  This is an
easy way to compute luminosity and mass-to-light ratio profiles, as
shown for A\,1689 in Paper~I. The main difficulty is the determination
of the background contamination which is estimated from the average
density far from the cluster. The measured values also strongly depend
on the contamination from a few very bright foreground galaxies which
appear in some cluster fields.  To estimate the total luminosity of the
cluster, it is necessary to correct for the magnitude cuts of the ``bright
galaxy'' catalog.  Assuming that the cluster galaxies follow a Schechter
luminosity function, the correction factor, defined explicitly in Paper~I,
ranges from 10 to 30 percent, depending on the filter and depth of
the individual images (for example in A\,1689 the correction factor is
1.28 in $B$, 1.11 in $R$, 1.27 in $I$).  The ``total'' luminosity of the
clusters is then computed in a radius defined by the parameter $r_{200}$
of the best fit model from \texttt{McAdam} (Sect.~\ref{sec:wlmass} and
Table~\ref{tab:mass_bestfit}). We refer to this estimate of the total
luminosity as $L_{R}^\mathrm{tot}$ in the rest of the paper.

We also use the ``elliptical galaxy'' catalogs, which we expect to be
cleaner from non-cluster contamination, especially in the bright magnitude
bins. Another advantage is that these catalogs are not limited at faint
magnitudes and thus include most of the elliptical cluster members up to
much fainter magnitudes than the ``bright galaxy'' catalogs. Therefore,
no incompleteness factor is necessary in the integration. The correction
for field contamination, again estimated from the average galaxy density
outside the cluster and based on the same catalogs, is less dominated by
bright galaxies than in the previous determination.  Again, the total
luminosity of the ellipticals is computed inside the virial radius
$r_{200}$ and is referred to as $L_{R}^\mathrm{ell}$ in the rest of
the paper.

In practice, the field contamination from a few bright galaxies remains an
issue for the total luminosity computed from the ``bright'' galaxies. We
therefore use preferentially the luminosity of the elliptical galaxies of
the clusters, which yields cleaner light-density maps with a more uniform
background distribution. We check that the two luminosities do not differ
strongly: the average ratio $L_{R}^\mathrm{tot} / L_{R}^\mathrm{ell}$
is 1.34, meaning that about 35\% of the cluster luminosity are missed
when only elliptical galaxies are considered.  Of course, this ratio
varies from cluster to cluster and is also sensitive to differences
in the galaxy populations. Using $L_{R}^\mathrm{ell}$ only means that
we assume that elliptical galaxies are best suited for tracing the
galaxy content of the clusters and its relation to the dark-matter
distribution. Support for this assumption is provided by several
studies showing a strong correlation between the light of early-type
galaxies and the mass distribution derived from weak-lensing analyses
\citep{smail97,clowe98,gray02,gavazzi04}.

\section{Two-dimensional mass maps}
\label{sec:2d-mass-maps}
\begin{table*}
  \caption{Global properties of the clusters in the sample: peak
    surface-mass density in terms of the standard deviation $\sigma$
    of the background fluctuations (col.~2); morphological
    characterization of the mass maps (col.~3); flag assessing the
    correlation between the mass maps and the light maps traced by the
    cluster ellipticals (col.~4); X-ray morphology as determined by
    \citet{smith05} (col.~5); ``overall classification'' assigned by
    \citet{smith05} based on HST lens modeling: for a cluster to be
    classified as ``regular'' it must contain a single mass component
    that is centered on the BCG, and the $K$-band luminosity of the BCG
    must contribute at least 50\% of the cluster's central $K$-band
    luminosity relative to the total cluster luminosity (col.~6).}
  \centering
  \begin{tabular}{lccccc}
    \hline\hline\noalign{\smallskip}
    Cluster &$\nu_{\mathrm{peak}}$ &  Mass & $M$ traces $L$? &
    X-ray & Overall \\
    & & morphology & & morphology & classification \\
    \noalign{\smallskip}\hline\noalign{\smallskip}
    A\,68   & 5.5 & Circular  & Y & Irregular  & Unrelaxed \\
    A\,209  & 4.9 & Elongated & N & Irregular  & Unrelaxed \\
    A\,267  & 3.6 & Elongated & N & Elliptical & Unrelaxed \\
    A\,383  & 3.6 & Circular  & Y & Circular   & Relaxed   \\
    A\,963  & 4.2 & Circular  & Y & Elliptical & Relaxed   \\
    A\,1689 & 9.6 & Circular  & Y & ---        & ---       \\
    A\,1763 & 7.4 & Elongated & Y & Irregular  & Unrelaxed \\
    A\,1835 & 8.7 & Circular  & Y & Circular   & Relaxed   \\
    A\,2218 & 6.5 & Elongated & Y & Irregular  & Unrelaxed \\
    A\,2219 & 5.0 & Circular  & Y & Irregular  & Unrelaxed \\
    A\,2390 & 7.4 & Elongated & N & ---        & ---       \\
    \noalign{\smallskip}\hline
  \end{tabular}
  \label{tab:morpho}
\end{table*}

We use an entropy-regularized maximum-likelihood technique, implemented
in \textsc{LensEnt2} \citep{marshall02}, to obtain mass reconstructions
for each cluster based on the ``faint galaxy'' catalogs in the $R$ band
after PSF correction.  Using the magnitude-selected ``faint galaxy''
catalog instead of the color-and-magnitude-selected ``red galaxy'' catalog
does not change the global shape of the derived mass distribution. To
avoid building under-dense regions with negative mass densities a mean
background of $100 M_{\sun}\,\mathrm{pc^{-2}}$ is artificially added
during the reconstruction. Since the mass distribution of clusters is
extended, the individual values of the mass density as reconstructed
with \textsc{LensEnt2} are spatially correlated. The effective scale
of this correlation is controlled by the \emph{Intrinsic Correlation
Function} (ICF) of the model which thus sets the resolution at which
mass structures can be detected. We use an ICF of 180\arcsec\ ($\sim
600\,h_{70}^{-1}\,\mathrm{kpc}$ at the cluster redshift) which represents
a good compromise between detail and smoothness of the mass maps.

Figs.~\ref{fig:lensent2} and~\ref{fig:lensent2b} show the results of
the two-dimensional mass reconstructions around each cluster. In each
plot the significance increases by $1\sigma$ between adjacent contours,
with the lowest level representing a significance of $2\sigma$ above
the mean background. For each cluster, $\sigma$ is the average level
of the noise peaks above the background, obtained from a randomization
over the orientation of the ``faint galaxy'' catalogs used in the
\textsc{LensEnt2} mass reconstruction. 200 such random catalogs are
produced for each cluster for the statistical analysis (see Paper~I).
The average value of $\sigma$ for the entire sample is $\sigma_\mathrm{av}
\simeq 80 M_{\sun}\,\mathrm{pc^{-2}}$.

\begin{figure*}
  \centering
  \includegraphics[width=0.42\textwidth]{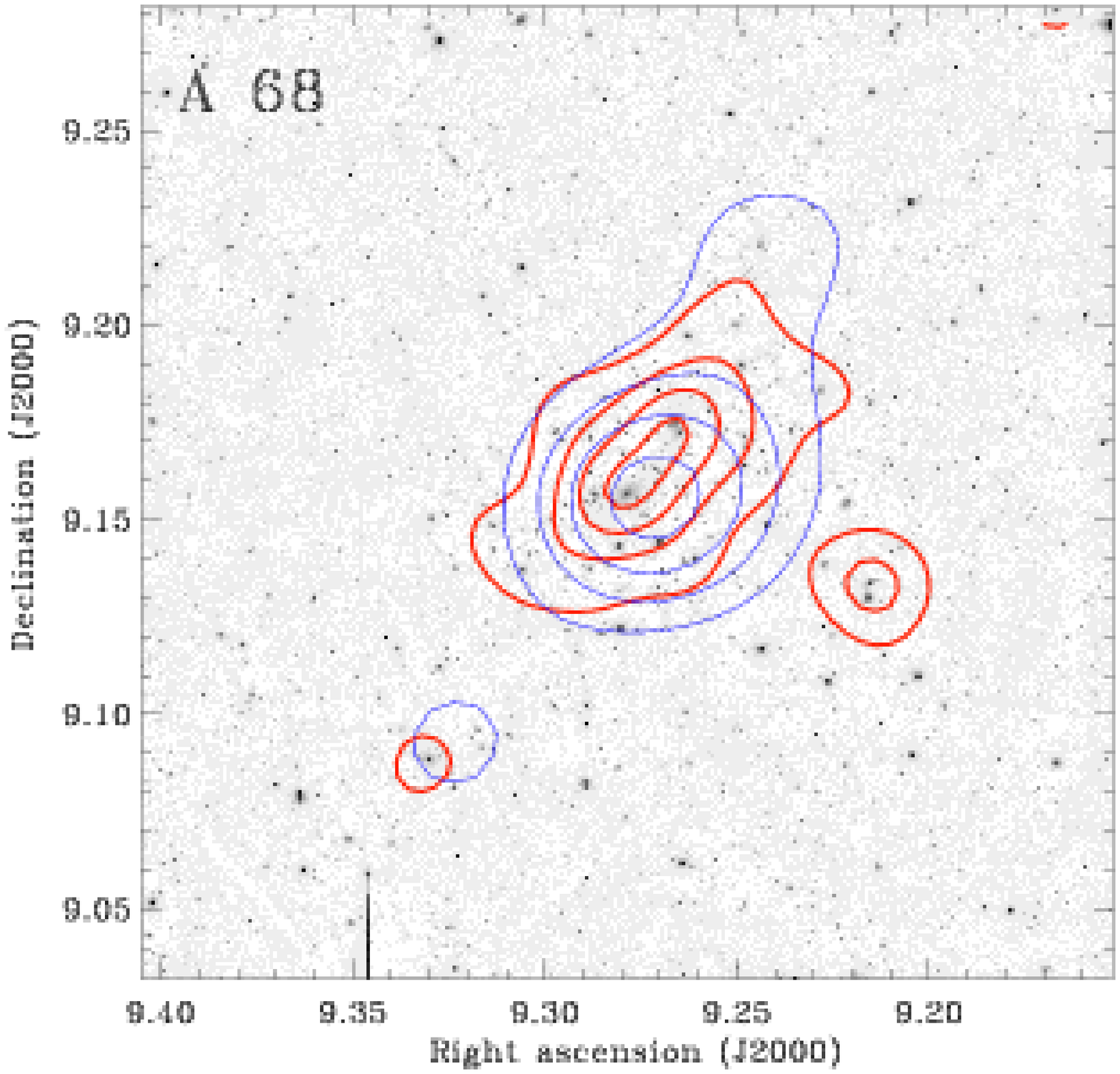}
  \includegraphics[width=0.42\textwidth]{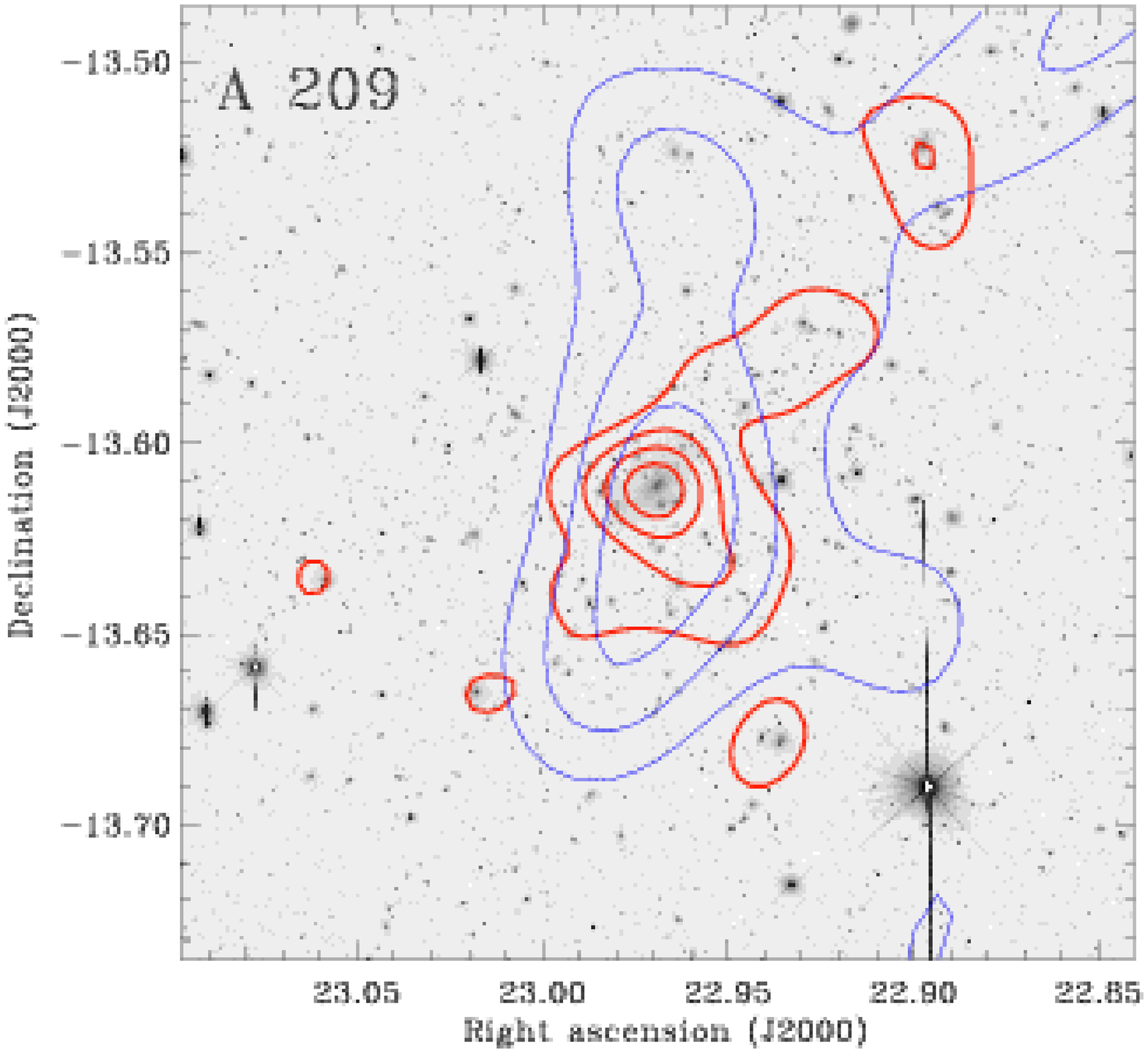}
  \includegraphics[width=0.42\textwidth]{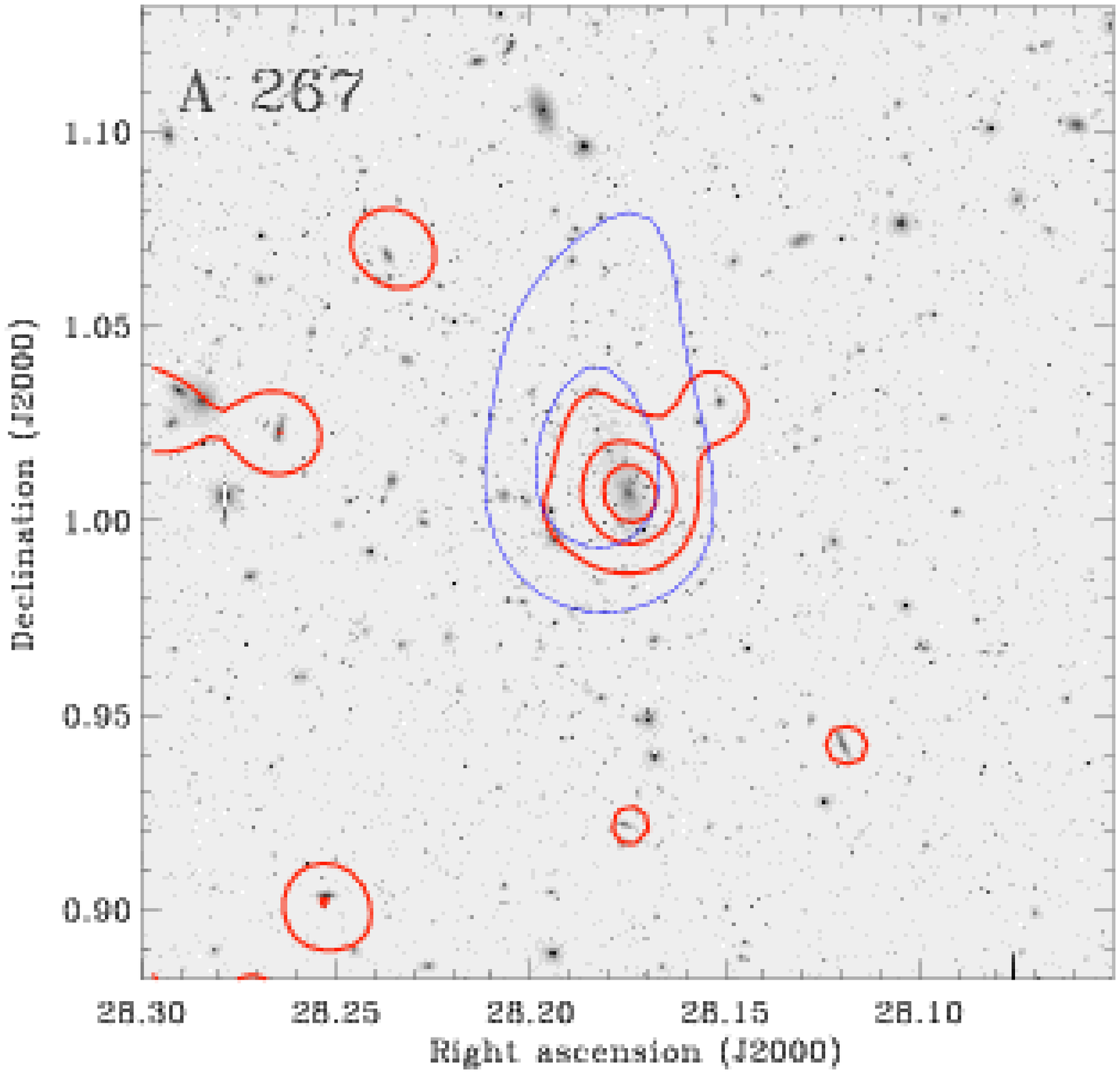}
  \includegraphics[width=0.42\textwidth]{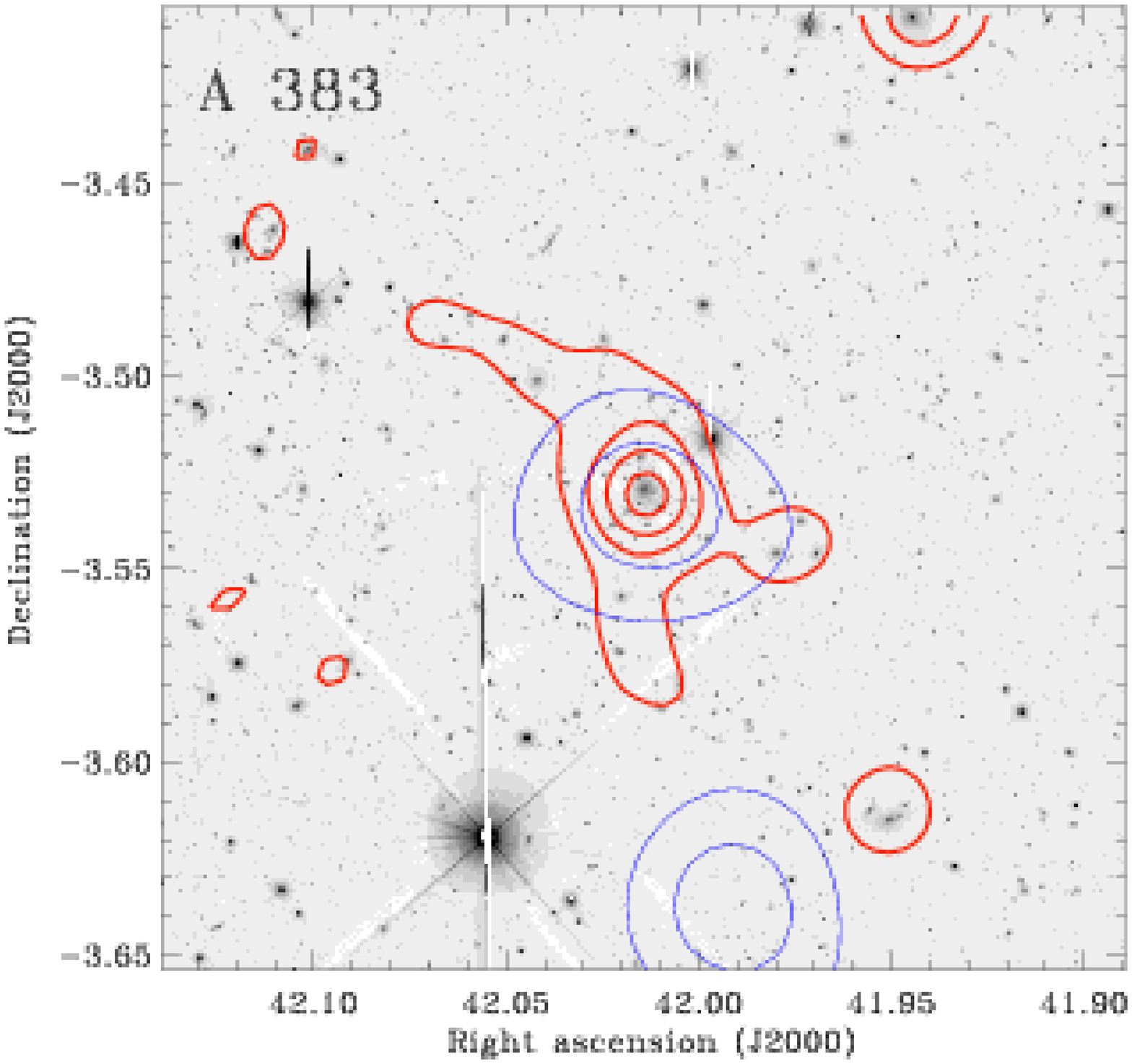}
  \includegraphics[width=0.42\textwidth]{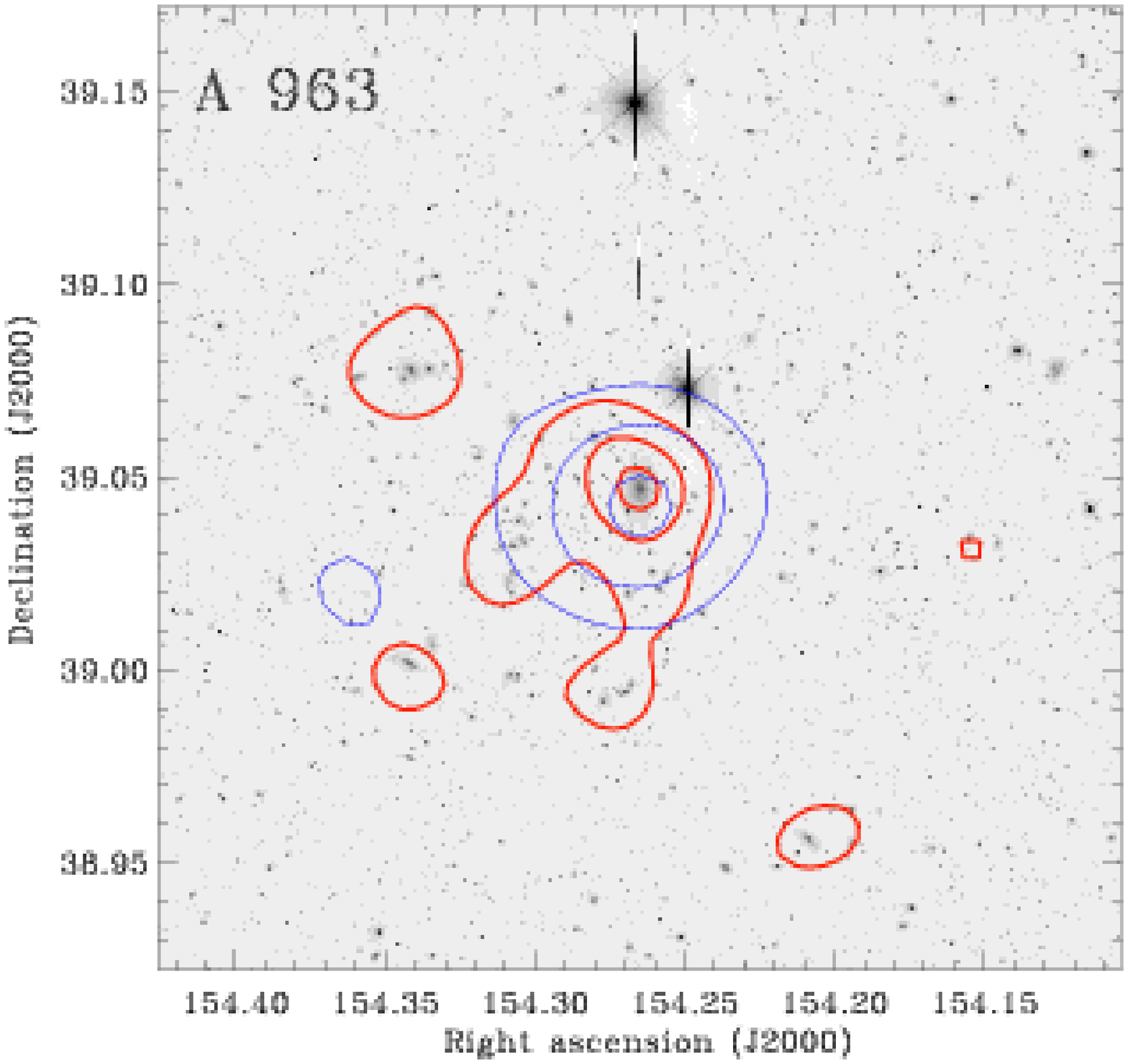}
  \includegraphics[width=0.42\textwidth]{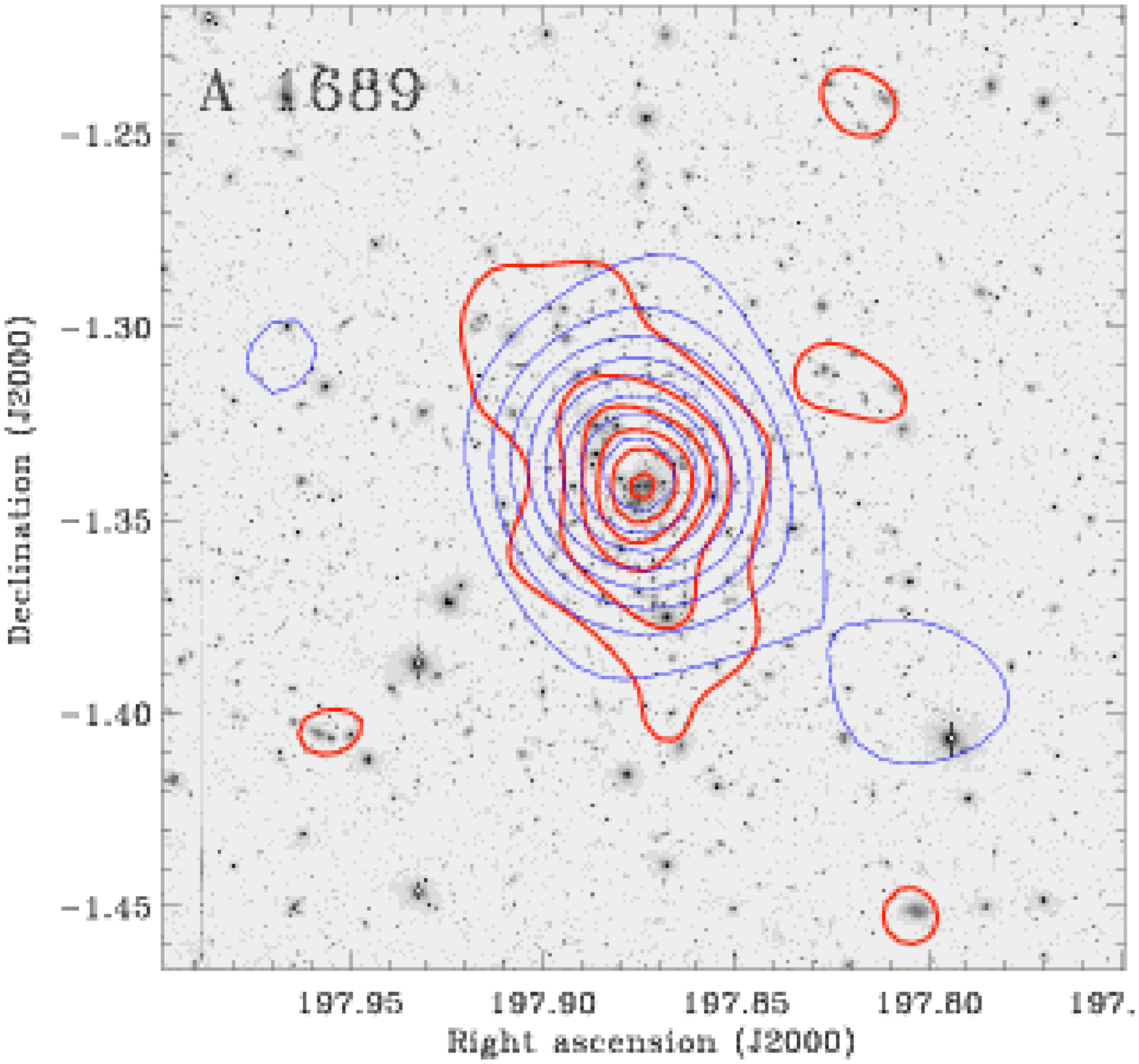}
  \caption{$15\arcmin \times 15\arcmin$ CFH12k $R$-band images of the
    eleven clusters of the sample. Thick contours represent the light
    density of ``elliptical'' galaxies selected in the $R-I$ versus
    $R$ color-magnitude diagram; thin contours represent the
    mass-density reconstruction from \textsc{LensEnt2} (see text for
    details).  The light contour levels are adjusted for each cluster
    according to its richness (or central density) and range from $2$
    to $3\times10^{5}\,L_{\odot}\,\mathrm{arcmin}^{-2}$. Slight
    offsets between the positions of the mass peaks and the locations
    of the BCGs are comparable to the resolution of the mass
    reconstruction of about 3\arcmin and thus not significant. North
    is up and East is to the left.}
  \label{fig:lensent2}
\end{figure*}

\begin{figure*}
  \centering
  \includegraphics[width=0.42\textwidth]{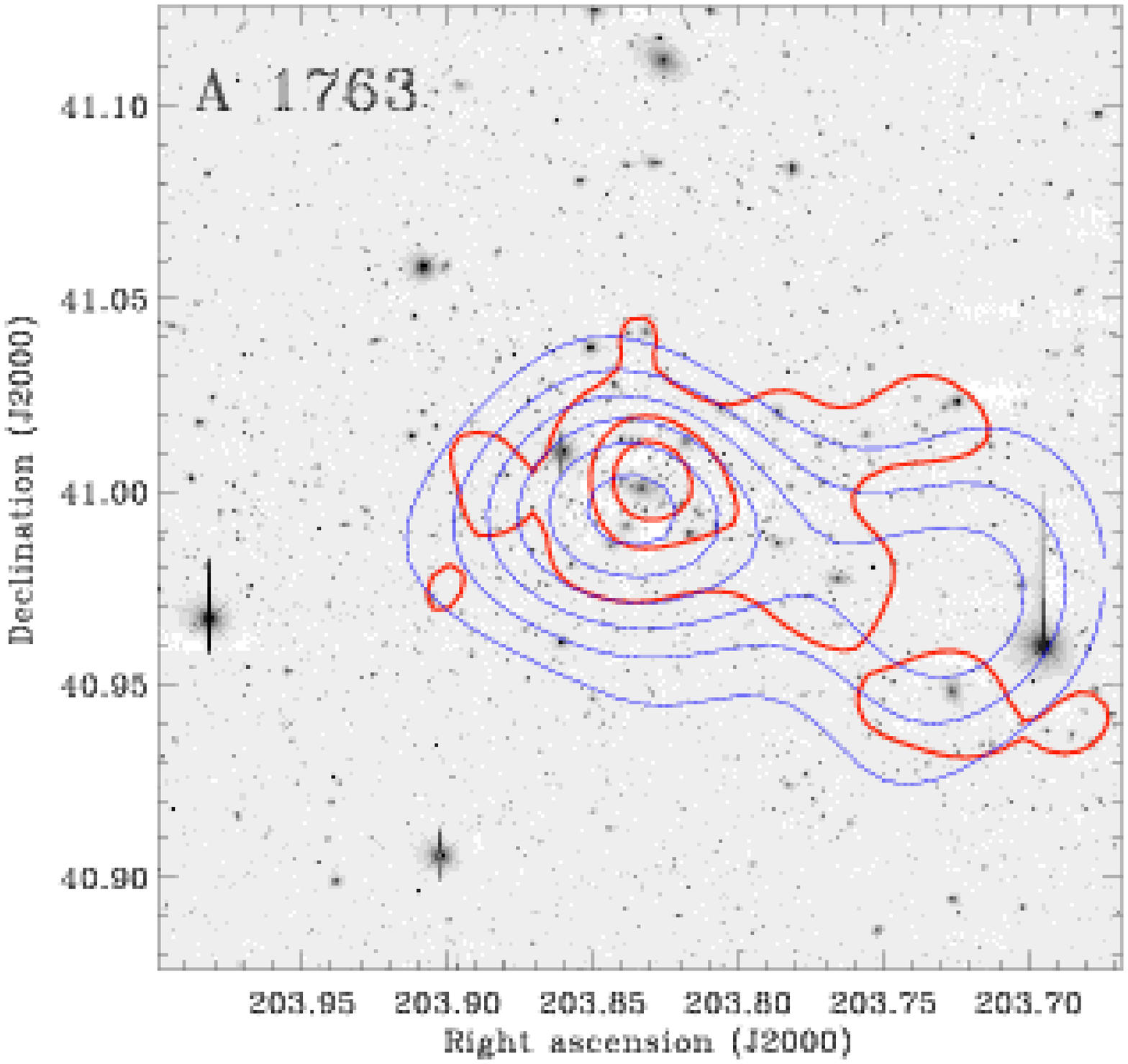}
  \includegraphics[width=0.42\textwidth]{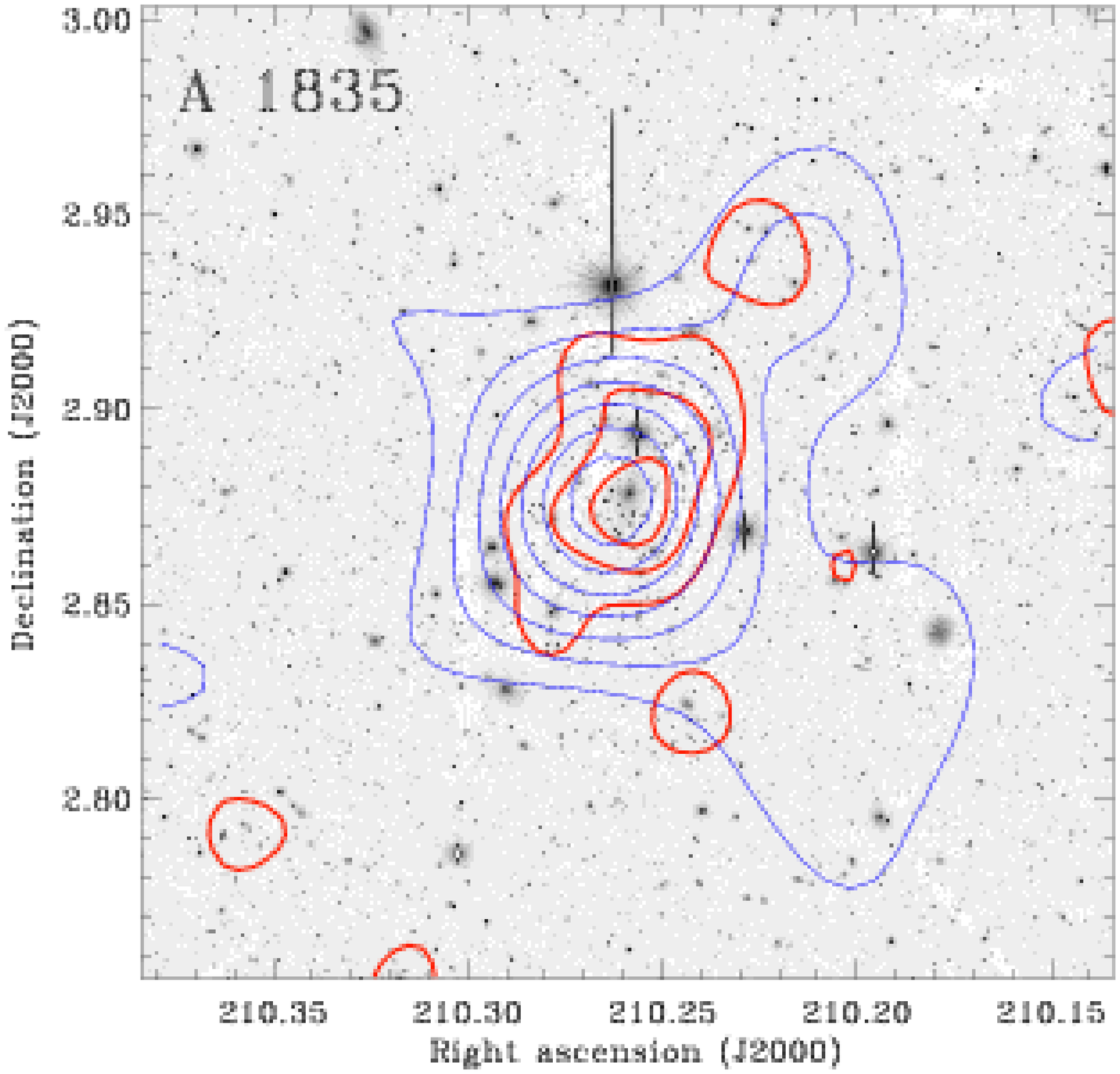}
  \includegraphics[width=0.42\textwidth]{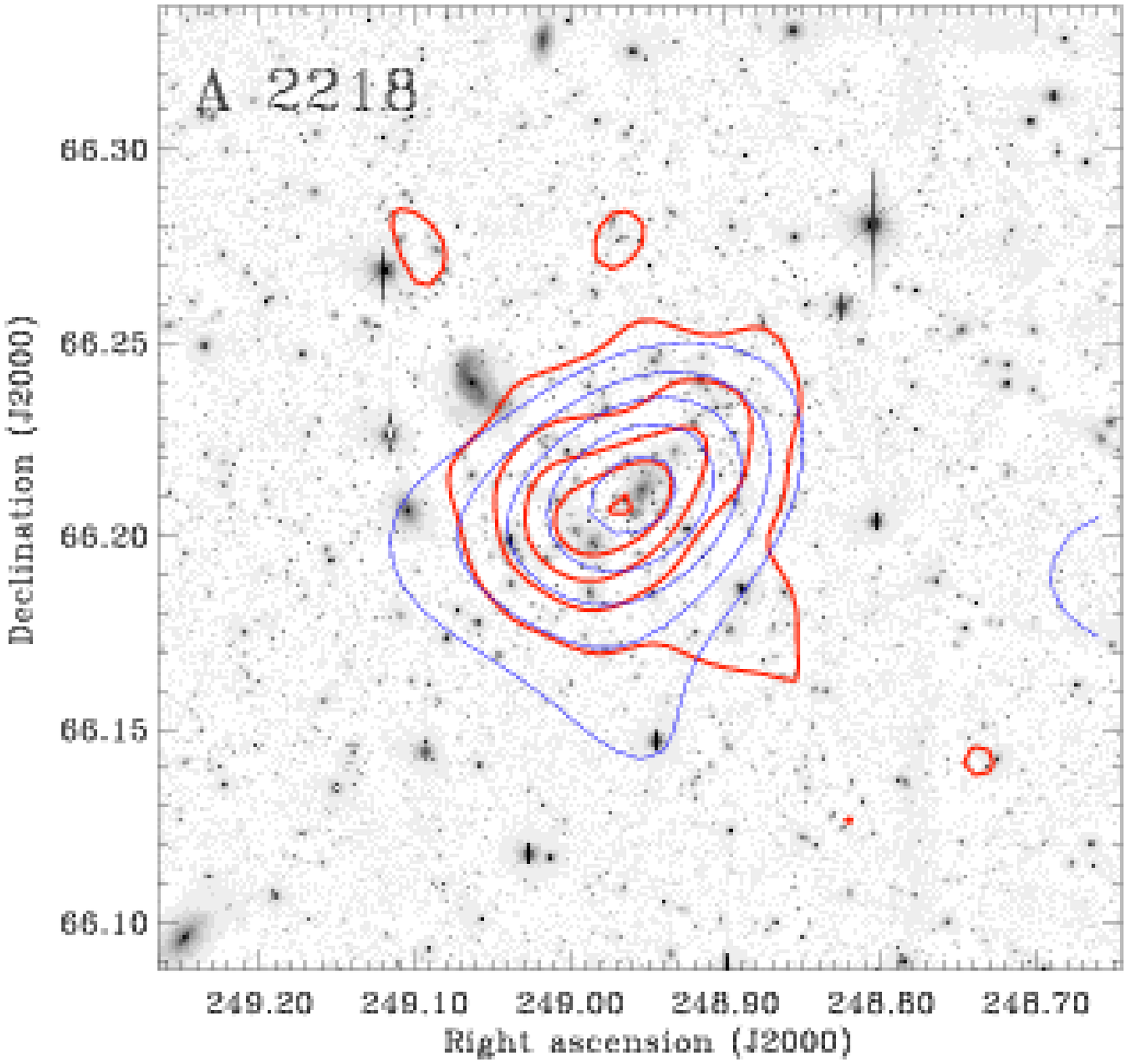}
  \includegraphics[width=0.42\textwidth]{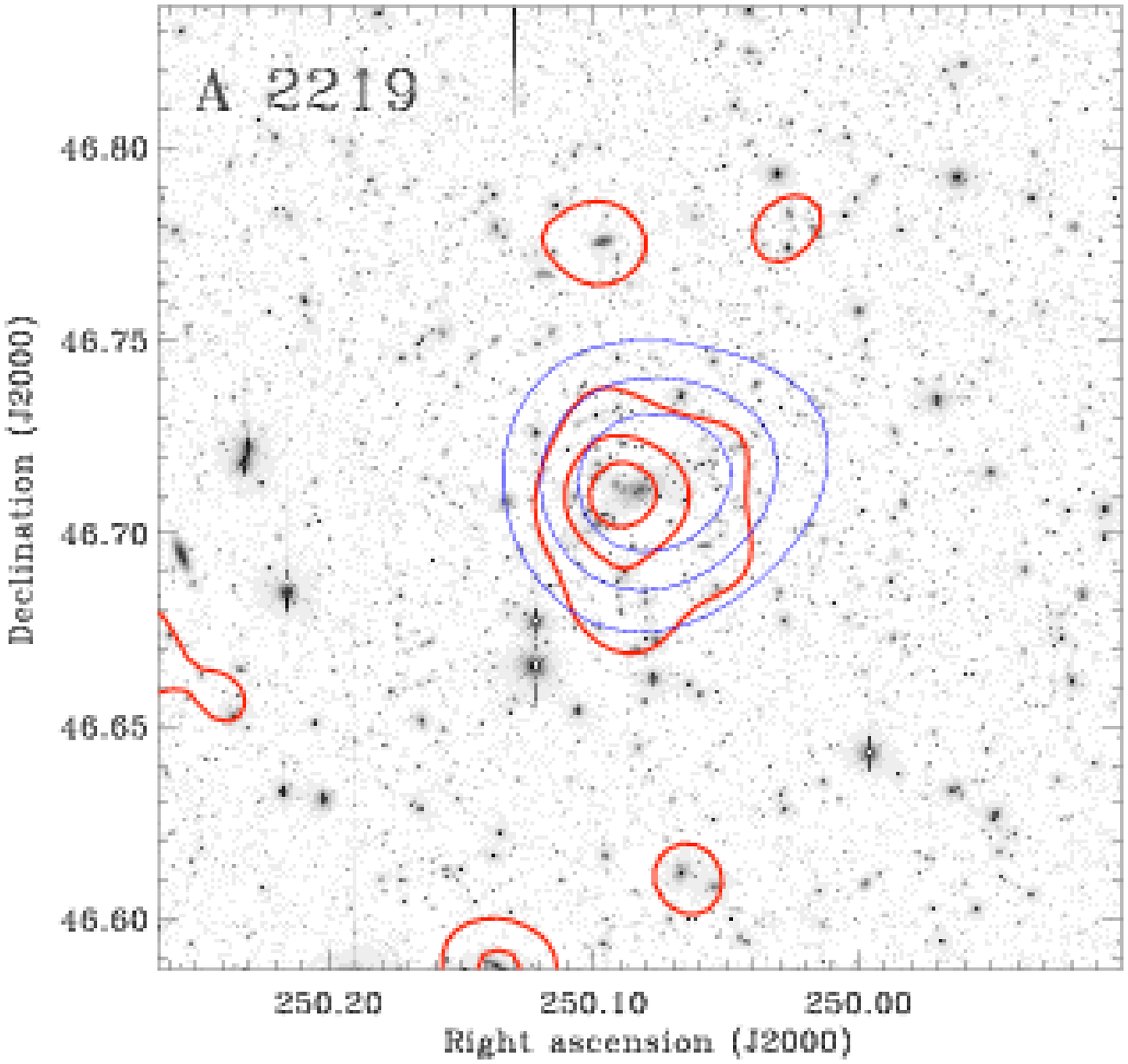}
  \includegraphics[width=0.42\textwidth]{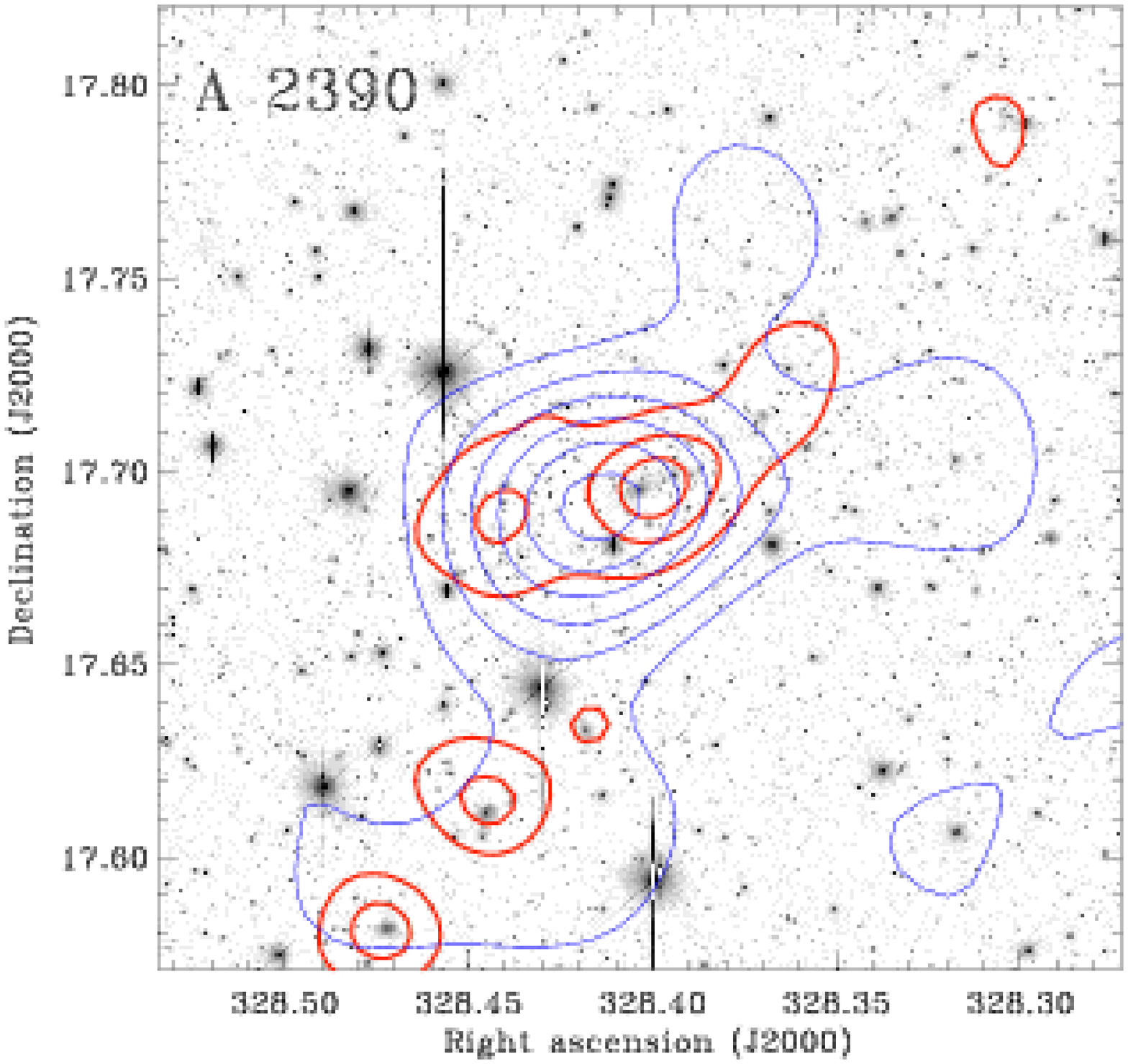}
  \caption{Continued from Fig.~\ref{fig:lensent2}.}
  \label{fig:lensent2b}
\end{figure*}

In all cases, the target cluster is detected at high significance.
Table~\ref{tab:morpho} gives the significance of the detection
$\nu_{\mathrm{peak}}$ in units of $\sigma$ as defined above. Even for
the least significant detection, A\,267, the main peak is detected
at about $3.6\sigma$. A quantitative assessment of the reality of the
mass clumps detected with \textsc{LensEnt2} outside the main cluster
component and at much lower nominal significance would require numerical
simulations that are beyond the scope of this paper. We here use the
mass maps solely to assess qualitatively the morphology of the cluster
mass distributions (Table~\ref{tab:morpho}). We crudely classify the
clusters as \emph{circular} or \emph{elongated} if the mass distribution
within the $3\sigma_\mathrm{av}$ contour has an ellipticity $(1-b/a)$
smaller or larger than $0.2$, respectively.

Also shown in Figs.~\ref{fig:lensent2} and~\ref{fig:lensent2b} are
contours of the light density of the ``elliptical'' galaxies as defined
in Sect.~\ref{ssec:catalogues}.  In Table~\ref{tab:morpho} we indicate
whether there is a good visual correlation between the mass and light
distributions.  Although these are rather qualitative criteria, most of
the ``elongated'' clusters have both a mass and light distribution that
is clearly not spherical.

Globally there is good agreement between the morphological information
from our X-ray, strong- and weak-lensing analyses, as well as the
distribution of light in elliptical galaxies. We also find good agreement
with the overall classification of \citet{smith05}: the four clusters
A\,383, A\,963, A\,1689 and A\,1835 correspond to the more relaxed and
spherical clusters of the sample, while A\,2219 and A\,68 are also close
to this category.

\section{Weak-lensing masses}
\label{sec:wlmass}

\subsection{Selection of background galaxies}
\label{ssec:background_galaxies}
One of the main difficulties in obtaining reliable weak-lensing mass
estimates is to ensure that only background galaxies are used in the
analysis. The catalogs must be as free as possible from contamination from
foreground or cluster galaxies to avoid attenuation of the weak-lensing
signal, averaged in radial bins or locally, from galaxies with purely
random orientations. The potential for such attenuation is largest
near the cluster cores where the galaxy density is highest, causing
the weak-lensing profiles to be flattened and the total masses to be
underestimated. For a quantitative test of this effect, we create mock
catalogs with the same density of sources as observed, distributed with
random orientations and a Gaussian ellipticity distribution similar to the
one observed after the PSF correction \citep{bardeau05}. These catalogs
are then ``lensed'' by a cluster with an NFW mass profile, and average
shear profiles are built in the same manner as for the observed catalogs
(see below).  The shear profiles are fitted by several mass profiles
and the total cluster masses estimated in this way compared to the input
cluster mass. To test the aforementioned attenuation effect contamination
from cluster galaxies is added to the lensed catalogs, with a number
density profile mimicking that of A\,1689. The resulting mass profiles
and total masses are again compared to the initial inputs. We find that
even a 10\%\ contamination by cluster members in the catalogs can cause
the reconstructed total mass to be underestimated by up to 20 to 30\%.

\subsection{Fit of the shear profile}
\label{ssec:shear_profile_fit}
Two parametric mass profiles are used to fit the lensing data: a
singular isothermal sphere (SIS) and an NFW profile \citep{navarro97}.
The former is characterized by a single parameter, the Einstein
radius $\theta_\mathrm{E}$ or, equivalently, the velocity dispersion
$\sigma_\mathrm{v}$, while the latter is described by two independent
parameters usually chosen as the scale radius $r_\mathrm{s}$ and the
concentration parameter $c= r_{200}/r_\mathrm{s}$.  The virial mass
$M_{200}$ is computed as a function of the virial radius $r_{200}$ and
the critical density $\rho_\mathrm{c}$ of the Universe at the cluster
redshift:
\begin{equation}
  \label{eq:M200-def}
  M_{200} = \frac{4}{3} \pi r_{200}^3 \times 200\,\rho_\mathrm{c} (z)
\end{equation}

\begin{table*}
  \begin{center}
    \caption{Results of the mass-profile fits, obtained with
      \texttt{McAdam} and using the ``red galaxy'' catalogs only, for
      both the NFW and the SIS profiles. For all fits the cluster
      center is assumed to coincide with the location of the brightest
      cluster galaxy.  The Einstein radius is computed at the average
      redshift of the background sources for each cluster (see text
      for details).  $M_{200}$ (also referred to as the virial mass)
      is the total mass included in a sphere of radius $r_{200}$.
      Also listed are the $R$-band luminosities of the clusters:
      $L_R^\mathrm{tot}$ is integrated within the ``bright galaxy''
      catalog and corrected for incompleteness, while
      $L_R^\mathrm{ell}$ is the total luminosity of the ``elliptical
      galaxies'' selected within the color-magnitude sequence $R-I$
      versus $R$. All luminosities are integrated within the virial
      radius $r_{200}$ and error bars are based on the uncertainty of
      the $r_{200}$ determination. }
  \label{tab:mass_bestfit}
    \begin{tabular}{l|ccc|cc|cc}
      \hline\hline
      Cluster  \raisebox{0ex}[2.5ex][0ex]{} 
      & $c$ & $r_{200}$ & $M_{200}$ & 
      $\sigma_{\mathrm{shear}}$ & $\theta_{\mathrm{E}}$ &
      $L_R^{\mathrm{tot}}$ & $L_R^{\mathrm{ell}}$ \\
      \raisebox{0ex}[2ex][1ex] & & ($h_{70}^{-1}$ Mpc) & ($10^{12}\,h_{70}^{-1}\,M_{\sun}$) & 
      (km/s) & $ (\arcsec)$ & ($10^{12}\,h_{70}^{-2}\,L_{\sun}$) &
      ($10^{12}\,h_{70}^{-2}\,L_{\sun}$) \\
      \hline
      A\,68   \raisebox{0ex}[2.5ex][0ex]{} 
      & $3.84 \pm 1.13$ & $1.49 \pm 0.18$ & $ 620 \pm 197$ &
      $ 880 \pm  65$ & $12.9 \pm 1.9$  & 
      $ 6.6 \pm 0.3$ & $ 5.5 \pm 0.3$ \\
      A\,209  
      & $3.00 \pm 0.92$ & $1.57 \pm 0.17$ & $ 719 \pm 204$ &
      $ 813 \pm  70$ & $12.4 \pm 2.1$ & 
      $ 8.8 \pm 0.7$ & $ 7.6 \pm 0.6$ \\
      A\,267  
      & $4.54 \pm 2.01$ & $1.15 \pm 0.23$ &  $272 \pm 146$ &   
      $ 634 \pm 116$ & $ 6.7 \pm 2.4$ & 
      $ 5.3 \pm 2.0$ & $ 4.6 \pm 1.7$ \\
      A\,383  
      & $2.62 \pm 0.69$ & $1.32 \pm 0.17$ & $ 419 \pm 146$ &
      $ 619 \pm  72$ & $ 7.4 \pm 1.7$ & 
      $ 7.0 \pm 2.5$ & $ 4.9 \pm 1.4$ \\
      A\,963  
      & $8.35 \pm 1.25$ & $1.33 \pm 0.10$ & $ 396 \pm 90$ &
      $ 812 \pm  67$ & $12.3 \pm 2.0$ & 
      $ 5.9 \pm 0.8$ & $ 4.2 \pm 0.5$  \\
      A\,1689 
      & $4.28 \pm 0.82$ & $2.25 \pm 0.14$ & $1971 \pm 336$ &
      $1277 \pm  37$ & $31.8 \pm 1.8$ & 
      $12.8 \pm 0.5$ & $11.0 \pm 0.4$  \\
      A\,1763 
      & $2.63 \pm 0.63$ & $1.93 \pm 0.14$ & $1386 \pm 263$ &
      $ 932 \pm  60$ & $15.0 \pm 1.9$ & 
      $12.9 \pm 0.8$ & $10.0 \pm 0.6$ \\
      A\,1835 
      & $2.58 \pm 0.48$ & $2.39 \pm 0.14$ & $2707 \pm 414$ &
      $1240 \pm  47$ & $26.6 \pm 2.0$ & 
      $16.8 \pm 0.8$ & $12.1 \pm 0.9$ \\
      A\,2218 
      & $6.86 \pm 1.30$ & $1.81 \pm 0.14$ & $ 971 \pm 215$ &
      $1040 \pm  50$ & $21.2 \pm 2.0$ & 
      $12.8 \pm 0.5$ & $ 6.8 \pm 0.3$ \\
      A\,2219 
      & $3.84 \pm 0.99$ & $2.25 \pm 0.18$ & $2094 \pm 435$ &
      $1175 \pm  53$ & $23.4 \pm 2.1$ & 
      $14.6 \pm 0.7$ & $11.1 \pm 0.8$ \\
      A\,2390   \raisebox{0ex}[2ex][1ex]
      & $5.26 \pm 1.43$ & $1.74 \pm 0.17$ & $ 943 \pm 246$ &
      $1015 \pm  54$ & $18.1 \pm 1.9$ & 
      $ 9.0 \pm 1.1$ & $ 6.6 \pm 0.8$ \\
      \hline
    \end{tabular}
  \end{center}
\end{table*}

In order to determine the cluster mass distribution from the weak-lensing
data we fit the shear pattern using \texttt{McAdam}, a Bayesian method
developed by \citet{marshall02} and \citet{marshall06}.  \texttt{McAdam}
works directly on the PSF-corrected faint-galaxy catalogs without any
radial binning of the data and is consequently more flexible than a fit
of a few data points obtained by averaging within circular annuli. The
output of \texttt{McAdam} is a probability distribution of the fitted
parameters, which is obtained using a maximum-likelihood estimator and a
MCMC iterative minimization.  We use \texttt{McAdam} on the ``red galaxy''
catalogs and fit the mass distribution of each cluster with a single
component, leaving the mass profile parameters ($\theta_\mathrm{E}$
for SIS, $c$ and $M_\mathrm{200}$ for NFW) as free parameters. A prior
on the concentration parameter $c$ is included with $2<c<10$, following
the results of N-body simulations of cosmological structure formation
at the galaxy cluster scale \citep{bullock01,hennawi07}.  As discussed
before, any residual contamination of the catalog by cluster galaxies, in
particular in the central area, may reduce the central shear signal and
thus flatten the deduced mass profile. As no strong-lensing information
is included, we add a prior on the lens center by assuming it to coincide
with the position of the brightest cluster galaxy (BCG).  This reduces
the number of free parameters in the global fitting but is of little
consequence for the final value of the total mass because the results of
the fits are dominated by the shear signal at large distance from the
cluster center. In addition to obtaining a global fit, we also compute
for each cluster the radial shear profile which allows a straightforward
assessment of the strength of the detected shear signal. The eleven
shear profiles derived from the ``red galaxy'' catalogs are shown
in Appendix~\ref{app:shear-profiles}.  A detailed cluster by cluster
comparison with previous mass estimates from independent weak lensing
studies is also included in this Appendix. In general, there is good
agreement between our measurements and previous ones, except in a few
cases like A\,1689 where there is a wide range of measured weak lensing
masses, using different methods. The most recent ones seem to converge to
a value close to the one presented in this paper. Note also that seven
out of our eleven clusters were imaged by \citet{dahle02} who fitted
their radial shear profiles with SIS profiles. Most of the clusters
have a best-fit value of the velocity dispersion $\sigma_{\mathrm{sh}}$
much higher than the present one, but their analysis was based on
shallower data taken in a much smaller field of view and so has much
larger systematic and random errors than our analysis.

The emphasis of the work presented here is on the total mass of the
clusters.  The resulting best fits as well as the internal errors
are summarized in Table~\ref{tab:mass_bestfit}. Given the correlated
nature of the parameters in our analysis and the weakness of some of
the detections, we focus our discussion on our estimates of the virial
radius $r_{200}$ and the virial mass $M_{200}$ which are robust results
of the weak-lensing analysis.  Combining strong- and weak-lensing effects
increases the accuracy of the mass reconstruction close to the center
and the fit of the concentration parameter $c$ of the NFW profile
\citep{broadhurst05b} but requires modifications of the likelihood
estimators used by \texttt{McAdam} in the central areas. Results from
a high-resolution study based on this approach will be presented in a
forthcoming paper (Hudelot et al., in preparation) and will also allow
us to discuss the distribution of values measured for the concentration
parameter $c$ across the sample as well as its cosmological consequences.

Fitting the shear profile with the SIS mass profile yields a
velocity dispersion $\sigma_\mathrm{shear}$ which can be converted
into an Einstein radius $\theta_\mathrm{E}$ using the value of
the ratio $D_{\mathrm{LS}}/D_{\mathrm{OS}}$ averaged over all the
sources with redshifts estimated from their photometric properties
\citep{bardeau05}. $\theta_\mathrm{E}$ does not depend strongly on
the exact value of $z_\mathrm{S}$ which turns out to be $z_\mathrm{S}
\simeq 1.1$ for all clusters.  We stress that errors due to differences
in the redshift of the background population from cluster to cluster
are negligible as we selected the clusters to lie in a narrow redshift
range so that the scaling factor $D_{\mathrm{LS}}/D_{\mathrm{OS}}$
varies by less than 5\% from cluster to cluster.A global error on the
mean redshift of the sources may introduce a systematic shift of the
scaling factor and hence the total mass, but its variation with $\langle
z_{\mathrm{S}}\rangle$ is so small that this cannot account for more
than a few percent, provided the sources are at redshift at least twice
the lens redshift.

\begin{figure}
  \centering
  \includegraphics[width=0.5\textwidth]{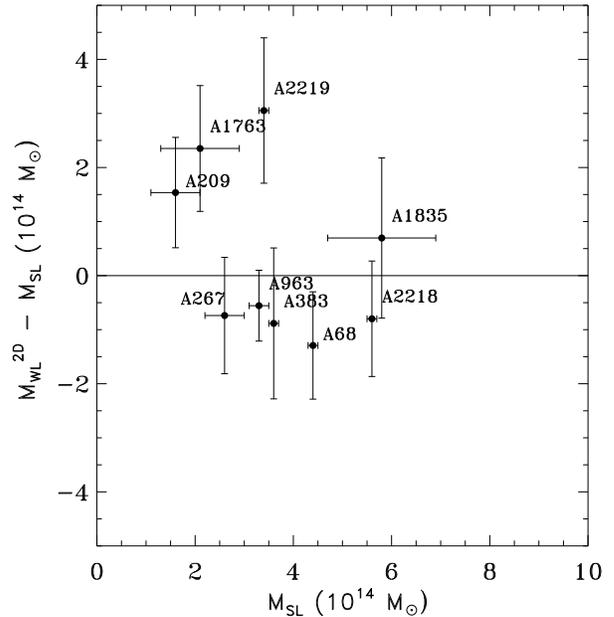}
  \caption{Projected mass enclosed within a radius of
    $500\,h_{50}^{-1}\,\mathrm{kpc}$ derived from the strong-lensing
    analysis of \citet{smith05} versus the difference between the
    weak- and strong-lensing masses measured in the same aperture.
    The weak-lensing masses have been converted to an Einstein-de
    Sitter cosmology with $H_0 = 50\,\mathrm{km\, s^{-1}\,Mpc^{-1}}$
    to allow a direct comparison with the strong-lensing masses.}
  \label{fig:SL_m200}
\end{figure}

A direct comparison between the masses from our weak-lensing
analysis and the masses measured by \citet{smith05} is
presented in Fig.~\ref{fig:SL_m200}. The strong-lensing ``total
mass'' is in fact the projected mass enclosed within a radius
$R=500\,h_{50}^{-1}\,\mathrm{kpc}$, extrapolated from the strong-lensing
modeling, whereas the virial mass $M_{200}$ is measured in a sphere
of radius $r_{200}$ which varies from cluster to cluster.  In order
to account for these differences we compute the projected mass inside
the projected radius $R=500\,h_{50}^{-1}\,\mathrm{kpc}$ using the NFW
parameters determined from the weak-lensing best fit, following the
relation
\begin{equation}
  \label{eq:M_WL-2D:general}
  M_{\mathrm{WL}}^{\mathrm{2D}} (<R) = 
  2 \pi r_{\mathrm{s}}^2 \,\Sigma_{\mathrm{cr}} \int_0^{R/r_{\mathrm{s}}}
  \kappa(y)\,y\,dy 
\end{equation}
with $r_\mathrm{s}\!=\!r_{200}/c$. The integral has been computed by
\citet{bartelmann96} and \citet{wright00} and the projected mass can
be rewritten in terms of the virial mass $M_{200}$ and the
concentration parameter $c$ as
\begin{equation}
  \label{eq:M_WL-2D:NFW}
  M_{\mathrm{WL}}^{\mathrm{2D}} (<R) = \frac{M_{200}}{\ln (1+c) -
    c/(1+c)}\, {\cal F} \left(\frac{R}{r_\mathrm{s}}\right)
\end{equation}
with ${\cal F} (x)$ defined as
\begin{equation}
  \label{eq:F-x}
  {\cal F} (x) = \left\{
    \begin{array}{ll}
      \displaystyle \frac{2}{\sqrt{1-x^2}} \artanh
      \sqrt{\frac{1-x}{1+x}}  
      + \ln \left(\frac{x}{2}\right) & \qquad (x<1) \\
      \vspace{.1cm}\\
      \displaystyle 1 + \ln \left(\frac{1}{2}\right) & \qquad (x=1) \\
      \displaystyle \frac{2}{\sqrt{x^2-1}} \arctan
      \sqrt{\frac{x-1}{x+1}} + \ln \left(\frac{x}{2}\right) & \qquad (x>1) \\
    \end{array}
  \right.
\end{equation}

The comparison between the two masses is shown in Fig.~\ref{fig:SL_m200}
and shows good agreement. No obvious bias appears in the sample although
there seems to be a dichotomy between two families of clusters. A\,209,
A\,1763 and A\,2219 have weak-lensing masses that appear to be
overestimated in comparison with those of the other clusters. By contrast,
the three clusters identified as ``relaxed'' clusters both in the weak-
and in the strong-lensing analysis (A\,383, A\,963 and A\,1835) show
good agreement between the two mass measurements.

\begin{figure}
  \centering
  \includegraphics[width=0.5\textwidth]{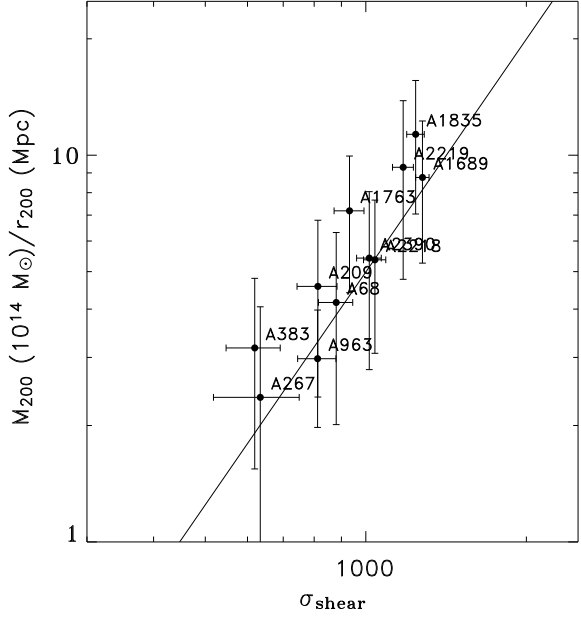}
  \caption{Comparison between the total mass $M_{200}$ divided by the
    virial radius $r_{200}$ and the velocity dispersion of the fitted
    SIS model. The straight line corresponds to $M_{200}/r_{200}
    \propto \sigma_{\mathrm{sh}}^2$ and confirms the correspondence
    between the two mass estimates (see text for details).}
  \label{fig:SIS_m200}
\end{figure}
In order to check the validity of our mass estimates, we also
compared masses deduced from the radial fits of the two different
parametric models (NFW and SIS). The total mass enclosed within the
virial radius is $M_{200}$ for the NFW profile and is proportional
to $\sigma_{\mathrm{shear}}^2 \times r_{200}$ for the SIS profile.
Fig.~\ref{fig:SIS_m200} compares these two quantities which show on
average the same behavior, so we are confident that using either the
virial mass or the mass deduced from the isothermal profile does not
change the validity of the scaling relations discussed below.  Note that
in both cases we use the 3D total mass enclosed within the virial radius
$r_{200}$.

\section{Global correlations}
\label{sec:optical}
With all measurements in hand we now investigate the correlations between
the lensing mass and immediate cluster observables like the optical
luminosity or X-ray characteristics such as luminosity and temperature.

\subsection{Correlations between mass and optical cluster properties}
\label{ssec:corr-mass-opt}
\begin{figure}
  \centering
  \resizebox{\hsize}{!}{\includegraphics{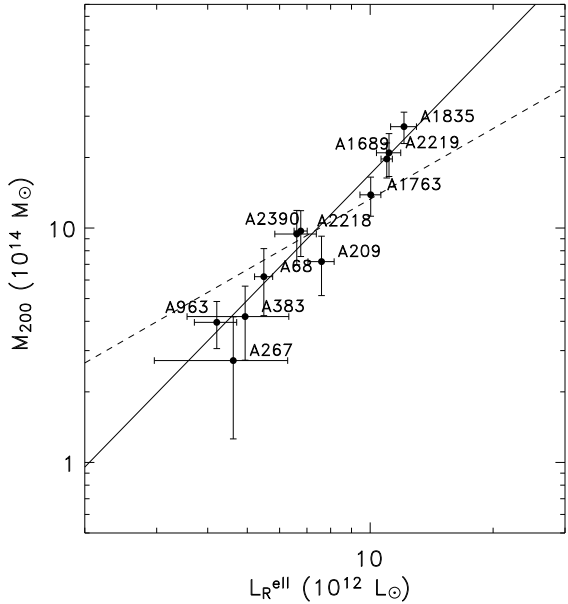}}
  \caption{Mass versus optical luminosity for the clusters in our
    sample. The mass is computed at the virial radius $r_{200}$
    derived from the best weak-lensing fits obtained with
    \texttt{McAdam}. The optical luminosity is computed in the
    $R$~band for the ``elliptical'' galaxies only, selected from the
    color-magnitude relation, and has been corrected for background
    contamination. Both quantities are projected (i.e.~integrated
    along the line of sight) and are therefore comparable. The dashed
    line represents a constant $M/L$ ratio of 133 in solar units while
    the solid line represents the best-fit power law $M \propto
    L^{1.8}$.}
  \label{fig:ML}
\end{figure}

Mass and luminosity are strongly correlated quantities as can be seen
in Fig.~\ref{fig:ML}. The $M/L$ ratio is usually representative of the
dynamical state of a cluster and is a tracer of its star-formation
history. One of the current issues of cluster research is whether
the $M/L$ ratio is a constant and universal value at least for rich
clusters, or whether it increases with mass, as recently suggested by,
e.g., \citet{popesso05}.

We here compute the $M/L$ ratio as the virial mass within the
virial radius $r_{200}$ ($M_{200}$) divided by the luminosity
$L_R^{\mathrm{ell}}$ as defined above. We use the virial mass instead of
the 2D projected mass $M_{200}^{\mathrm{2D}}$ because the latter depends
on the concentration parameter $c$, which is poorly constrained in our
study. We, however, convince ourselves that changing $c$ from 2 to 10
decreases the deduced projected mass by only less than 20\%. The optical
cluster luminosities are carefully corrected for background contamination
as well as Galactic extinction and are also $k$-corrected. We do not
apply any correction for luminosity evolution which is likely to be small
at the low redshift of our target clusters. To obtain a global $M/L$
ratio for all clusters we apply the average correction factor to the
total luminosity, 1.34. In practice, we find an average value for the
whole cluster sample before correction for the total luminosity
\begin{equation}
  \label{eq:ML_ell}
  \langle M/L_R^{\mathrm{ell}} \rangle = (133 \pm 50)\,h_{70}\, (M/L)_{\odot}
\end{equation}
or globally for all galaxies
\begin{equation}
  \label{eq:ML_tot}
  \langle M/L_R^{\mathrm{tot}} \rangle = (100 \pm 38)\,h_{70}\,(M/L)_{\odot}\,.
\end{equation}

These values are close to the value of
$M/L^{\mathrm{all}}_V=180^{+210}_{-110}\,h\,(M/L)_\odot$ obtained
by \citet{smail97} in their early weak-lensing study of a sample of
rich clusters.  On the other hand, \citet{carlberg96,carlberg97} find an
average value of $\langle M/L_r \rangle = (289 \pm 50)\,h_{100}\,(M/L)_{r,
\sun}$ for the $M/L$ ratio of rich clusters from the CNOC survey
(a sample of 14 clusters spanning a redshift range $[0.17-0.54]$).
Theirs are, however, global dynamical values which overestimate the
$M/L$ ratio measured at the virial radius $r_{200}$ by approximately
20\% \citep{carlberg97}. Although, all considered, their average $M/L$
ratio is still higher than the one found in our analysis, the discrepancy
is barely significant when all sources of error are taken into account.
For A\,2390, the only galaxy cluster in common between the two samples,
the velocity dispersion as well as the $M/L$ point toward a higher
mass than the weak lensing mass measured in this paper. However, their
determination of the virial radius differs significantly from ours. As
a consequence, the virial mass and the total luminosity are not measured
in the same area and the comparison is not conclusive.

If we drop the assumption of a constant $M/L$ ratio for clusters,
Fig.~\ref{fig:ML} shows a strong correlation between mass and light,
steeper than for a constant $M/L$, with massive clusters having higher
$M/L$ ratios. A log-log fit of the mass-luminosity relation, including
errors on the mass measurements shows a power law dependence with an
index $\alpha = 1.80 \pm 0.24$, or equivalently
\begin{equation}
  \label{eq:ML_vs_L}
  M/L \propto L^{0.80 \pm  0.24}\,.
\end{equation}
The tendency of increasing $M/L$ ratio with virial mass is significant,
even though the mass range of the present sample is rather limited.
Previous analyses have already shown similar trends, in particular
in a detailed comparison between dynamical mass estimates from the
SDSS and luminosities coming from either optical or X-ray measurements
\citep{popesso05}.  Their sample spans a wide mass range, from groups
up to massive clusters, and although there is a large dispersion of
their data points, the correlation between $M/L$ and mass is confirmed,
but with a lower slope than in the present study.  However, although
the mass range of their sample is wider than ours their dynamical mass
determination may not be as accurate as the weak lensing masses in the
present analysis.  Complementary lensing mass measurements for lower
mass clusters have been obtained by \citet{parker05} using the CNOC
galaxy groups.  They show some evidence for an increase in the $M/L$
ratio from poor to rich galaxy groups. This result is in agreement with
theoretical predictions obtained by comparing the Press-Schechter mass
function with the observed luminosity function \citep{marinoni02}. The
$M/L$ ratio in the high mass range, typical of rich clusters of galaxies,
scales as $L^{0.5 \pm 0.26}$, in close agreement with our observations.
However, \citet{marinoni02} assumed a rather high value of $\sigma_8=0.9$.
Following the latest WMAP3 analysis, a lower $\sigma_8=0.75$ might change
their $M-L_{\mathrm{opt}}$ relation by lowering the normalization but not
the power-law exponent \citep{reiprich06}.

This scaling behavior is predicted in semi-analytical models of
galaxy formation and has been interpreted as a decrease in galaxy
formation efficiency in rich and dense environments due to the long
cooling time of hot gas \citep{kauffmann99}.  Indeed, recent X-ray
observations have shown that in the centers of dense clusters the gas
cools less efficiently than predicted by the standard cooling flow
model \citep{Peterson01,Fabian03}. By injecting energy in the form of
radio jets, AGN may be responsible for switching off cooling at the
centers of massive haloes, thus preventing the formation of very bright
structures \citep{Bower06}. Also, star formation is quenched in newly
infalling galaxies through a variety of physical processes, some of
which, e.g.~ram pressure stripping of gas, are the more efficient the
denser the environment \citep[e.g.][]{Treu2003}.

\subsection{Correlation between optical and X-ray luminosities}
\label{ssec:corr-opt-X}
There is some correlation between the clusters' optical and X-ray
luminosities, although the luminosity range is not very extended in
our sample of bright X-ray clusters. \citet{popesso05} performed
linear fits between these quantities using a large sample of
low-redshift groups and clusters of galaxies from the RASS-SDSS Survey
over more than two decades in luminosities. For the correlation
between optical and X-ray luminosities, they find a best fit law: 
\begin{displaymath}
  \log ( L_{\mathrm{opt}} / 10^{12} L_\odot) = 0.64 \log
  (L_{\mathrm{X}} / 10^{44}\,\mathrm{erg\,s}^{-1}) + 0.45 \pm 0.15\,. 
\end{displaymath}
Instead of doing similar fits with our sample, which may have little
physical meaning, we simply compare the averaged optical luminosity
($L_R^\mathrm{tot} = 10.2 \times 10^{12}\,L_\odot$) at the mean X-ray
luminosity of the sample ($\langle L_{\mathrm{X}}\rangle = 10.5 \times
10^{44}\,\mathrm{erg\,s^{-1}}$).  This value corresponds closely to
the value predicted by the relation of \citet{popesso05}.  We therefore
confirm their normalization from an independent sample with great accuracy
and demonstrate that this normalization does not change out to $z \sim
0.2$ at high $L_{\mathrm{X}}$.

\subsection{Correlation between cluster mass and X-ray luminosity}
\label{ssec:corr-mass-LX}
\begin{figure*}
  \centering
  \includegraphics[width=0.48\textwidth]{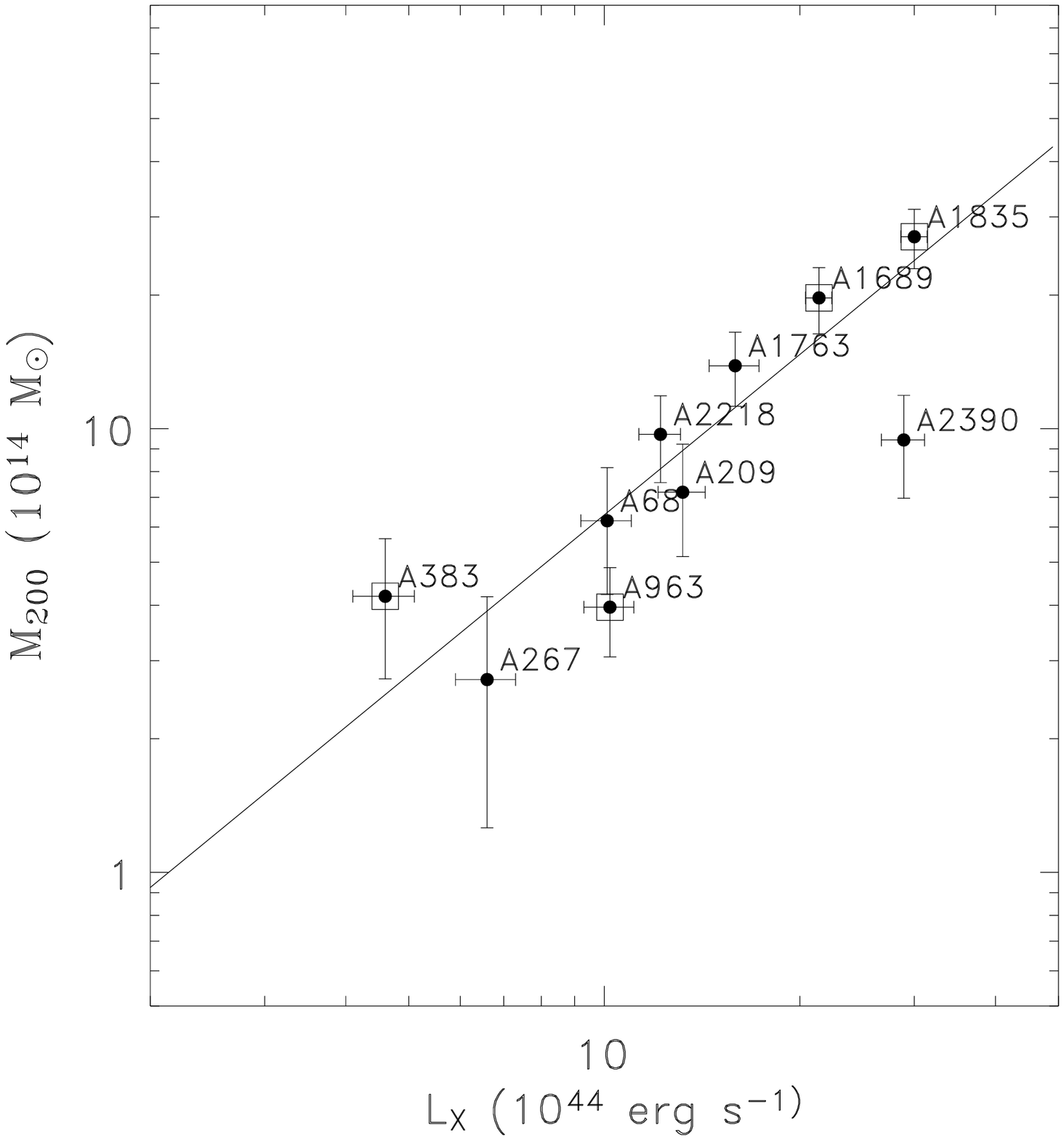}
  \includegraphics[width=0.48\textwidth]{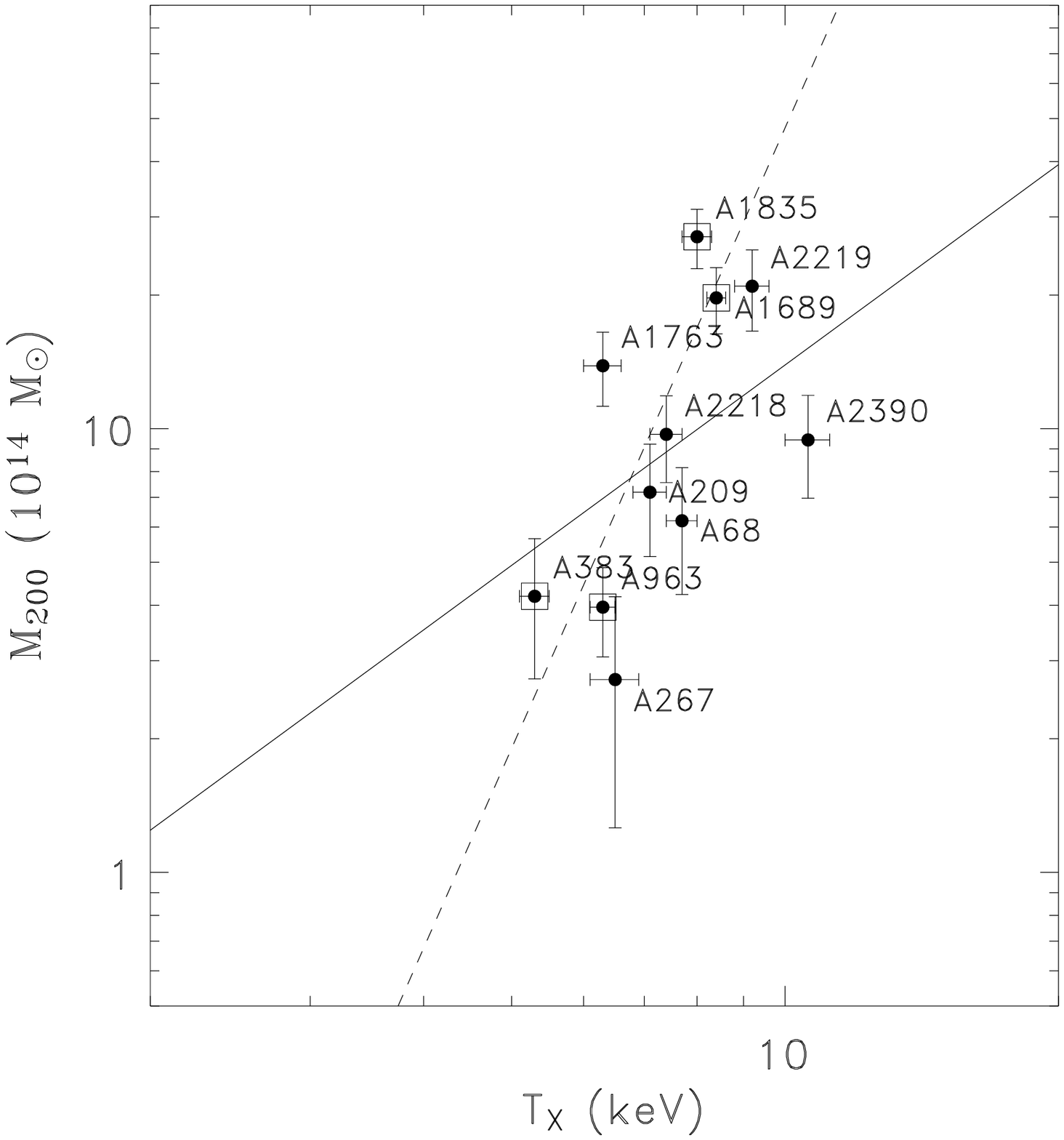}
  \caption{\textit{Left:} Weak lensing 3D virial mass $M_{200}$ versus
    X-ray luminosity. The best fit line has a slope $\alpha = 1.20 \pm
    0.16$ and is discussed in the text. \textit{Right:} Weak lensing
    3D virial mass $M_{200}$ versus X-ray temperature.  The straight
    line corresponds to a $M_{200} \propto T^{3/2}$ relation while the
    dashed line corresponds to the best fit power law relation $M 
    \propto T^{4.6 \pm 0.7}$. The
    virial mass $M_{200}$ is derived from the best weak lensing fits
    obtained with \texttt{McAdam}, temperatures are derived from XMM
    data \citep{zhang07}, completed by ASCA data from \citep{ota04}
    for A\,2219. In both plots, the 4 clusters with cooling core or
    relaxed properties are marked wit empty boxes.}
  \label{fig:MLxTx}
\end{figure*}

We now compare the deprojected mass derived from the weak lensing
analysis with the cluster X-ray properties. We consider the virial mass
$M_{200}$ enclosed within $r_{200}$, the result of the weak lensing best
fit performed with \texttt{McAdam}.  Bolometric X-ray luminosities are
measured from XMM-Newton data and integrated up to $2.5 r_{500}$, a radius
which is close to twice our virial radius $r_{200}$ \citep{zhang07}. They
are listed in Table~\ref{tab:sample}.  Fig.~\ref{fig:MLxTx} shows
the correlations between $M_{200}$ and the cluster X-ray properties.
A linear regression between mass and luminosity (including errors in
both axes in log-log space) gives the scaling relation
\begin{equation}
  \label{eq:M-LX}
  M_{200}/ 10^{12} M_\odot =  40^{+23}_{-15} \,\left(L_{\mathrm{X,bol}}
    / 10^{44}\,\mathrm{erg\,s^{-1}}\right)^{1.20 \pm 0.16}\;.
\end{equation}
As the slope is close to 1, one may also consider a constant mass to
(X-ray) light ratio for which we obtain an average value of
$M/L_{\mathrm{X}} = 67 \pm 24$ in units of ($10^{12} M_\odot /
10^{44}\,\mathrm{erg\,s^{-1}}$).  Inverting the $L_{\mathrm{X}}-M$
relation gives
\begin{equation} 
  \label{eq:LX-M}
  L_{\mathrm{X,bol}}/ 10^{44}\,\mathrm{erg\,s^{-1}} = 
  2.16 ^{+0.71}_{-0.53} \left(M_{200}/10^{14}\,M_\odot \right)^{0.83 \pm 0.11} 
\end{equation}
which seems in contradiction with some previous determinations of
the correlation between mass and X-ray luminosity where the slope
of the $L_{\mathrm{X}}-M$ relation is on the order of 1.3 to 2.0
\citep{reiprich02,popesso05,maughan07}. From the theoretical point of
view, a relation $L_{\mathrm{bol}} \propto M^{4/3}$ is expected in
a self-similar model for clusters, while including ``pre-heating''
in the physical processes of cluster formation leads to a relation
given by $L_{\mathrm{bol}} \propto M^{11/6}$ \citep{evrard91}. An
accurate determination of the observed $L_{\mathrm{X}}-M$ relation
is quite difficult with our present mass estimates, also because the
perturbation induced on the measurements by the central cooling core
of some clusters modifies the sample properties.  Note also that our
present normalization of the $L-M$ relation is only 60\% higher than the
one determined by \citet{maughan07}, contrary to the normalization found
by \citet{popesso05} which is lower than our value by a factor of 3 at
$10^{15} M_\odot$ but their $L-M$ relation presents a very different 
slope, close to 2 when using bolometric X-ray luminosities.
Going further in this analysis is presently difficult because of the
limited size and the small mass range of our sample. In addition,
our sample is X-ray flux-limited which may introduce some bias into
the relations involving $L_{\mathrm{X}}$, because for a given mass,
only the high luminosity clusters are selected \citep{reiprich06}.

\subsection{The mass-temperature relation}
\label{ssec:corr-mass-TX}
The X-ray temperatures have been measured on detailed XMM observations
\citep{zhang07}. The global values are the results of the volume averaged
radial temperature profiles between $0.2-0.5 r_{200}$. As shown in
Fig.~\ref{fig:MLxTx}, we presently have too large uncertainties and
dispersion in the measurements to correctly fit a power law for the
$M-T$ relation. Instead, we simply fix the slope to what is expected if
clusters are in hydrostatic equilibrium, $M \propto T^{3/2}$. Then we
find a normalization
\begin{equation}
  \label{eq:MTX}
  M_{200} / 10^{14} M_\odot = 0.44 ^{+0.39}_{-0.21} \,\left(
    \frac{T_{\mathrm{X}}}{1\,\mathrm{keV}} \right) ^{3/2}\,.
\end{equation}
This normalization deserves some comment because it is one of the few
attempts at fixing it with weak lensing mass measurements.
\citet{pedersen06} also provided a tentative determination of the
$M-T$ relation with weak lensing masses. From their whole sample of
rich clusters, spread over a large redshift range, they find
\begin{displaymath}
  M_{\mathrm{8\,keV}} = (0.78 \pm 0.14)\times 10^{15}\, h^{-1}\,M_\odot \,.
\end{displaymath}
The normalization translates to a value of $0.49 \pm 0.09$ in our
units (Eq.~\ref{eq:MTX}) and is in very good agreement with the
present work. Our value is also quite close to a recent prediction
obtained from a sample of simulated clusters realized with the most
recent implementation of the Tree-SPH code GADGET2 \citep{springel05}.
For the most massive clusters, \citet{ascasibar06} find a $M-T$
relation of
\begin{equation}
  \label{eq:MTX:Ascasibar}
  M_{200}/ 10^{14} M_\odot = 0.473 \, h_{70}^{-1} \,\left(
    \frac{T_{\mathrm{X}}}{1 \mathrm{keV}} \right) ^{3/2}\,,
\end{equation}
also in good agreement with our result, although their relation is
established at $z=0$. However, the authors claim that their
normalization is lower than the values found in previous simulations
and explain this partly by improvements of the treatment of entropy
conservation.

\subsection{Discussion}
\label{ssec:discussion}
The scatter in the $M-T$ and $L-T$ relations should be representative
of the diversity in the cluster histories but it is not easy to analyze
because our sample is small. The individual mass measurements still
have large uncertainties associated to the weak lensing method: masses
are underestimated if the catalogs are not completely cleaned from
foreground or cluster contamination and the low density of background
sources used for the weak lensing reconstruction adds another source
of noise which strongly limits the accuracy of the measurements. These
systematic uncertainties, which are difficult to quantify without detailed
simulations of mock catalogs, are presently not taken into account in
this analysis . The remaining uncertainty on the concentration parameter
adds at least another 20\% uncertainty on the total mass, included in
our present mass errors budget. All in all, weak lensing masses cannot
presently be determined with accuracy better than 40 to 50\%. However,
the tight correlation between mass and optical luminosity suggests that
some of these biases partly cancel when measuring the mass or the total
luminosity; this is the case for all geometric departures from spherical
symmetry (ellipticity of the light/mass distributions or projection
effects). Systematic biases in the mass determinations should not change
the slope of the scaling relations dramatically while some uncertainty
in their normalization remains. But since the slopes of the $L_X-M$,
$T_X-M$ and $L_ {opt}-M$ relations are all shallower compared to previous
results, we may also suspect some scale dependant biases, most probably at
the low mass end of our sample, where the weak lensing measures are the
most difficult to caracterize. Further improvements in the weak lensing
methodology are in progress and may help clarifying this possible
bias.

Moreover, although the sample was initially selected for its
homogeneity, at least in X-ray properties ($L_{\mathrm{X}} >
4\times10^{44}\,h_{70}^{-2}\,\mathrm{erg\,s^{-1}}$), at least three
out of the eleven clusters (i.e.~about 25\% of the sample) present a
strong central cooling core (namely A\,383, A\,963 and A\,1835) which
perturbs the total X-ray luminosity (clusters are overluminous) and
the X-ray temperature (clusters are too cool) for a given mass. This
effect is partly taken into account in the way \citet{zhang07} measured
the X-ray temperatures, excluding the X-ray signal in the central core
($r < 0.2 r_{500}$). Another five clusters show signs of non-sphericity
in the mass distribution. Following \citet{popesso05} we suspect that
most of the dispersion in the scaling relations is due to the intrinsic
dispersion on the X-ray properties of these rich clusters although
the present mass uncertainties add a significant fraction of the total
scatter. For the optical luminosity of clusters the link with the total
mass is tighter. This is an indication that early-type galaxies are good
tracers of the mass in clusters and that the dispersion in $M$ with $T_X$
is real.

\section{Conclusions}
\label{sec:conclusions}
We have presented the first weak-lensing analysis of a homogeneously
selected sample of eleven X-ray luminous clusters in a narrow redshift
slice at $z\sim0.2$. Using wide-field imaging in three bands ($B$, $R$
and $I$) covering up to 5\,Mpc in radius around the cluster targets we are
able to disentangle between foreground, cluster and background galaxies.
The weak lensing signal is always well detected up to 2\,Mpc, but
generally extends out to the edge of the field at lower signal-to-noise.

The weak lensing methodology used in the paper has already been tested and
validated on the cluster A\,1689 which yields the highest signal-to-noise
detection in our sample \citep{bardeau05}.  To measure galaxy shapes
and correct them for PSF anisotropy and circularization, we used the
\textsc{Im2shape} tool.  Our reliance on \textsc{Im2shape} is justified
by the results of the Shear Testing Program \citep[STEP,][]{heymans06}
which finds it to be a promising alternative to the popular KSB method
\citep{kaiser95}.  Cluster masses were computed using the \textsc{McAdam}
software \citep{marshall06} which performs a two-dimensional fit of
the individual galaxy shape information of the ``background'' galaxy
catalog.  We were thus able to determine: 1)~$M_{200}$ derived from
fitting NFW profiles (note that only weak constraints could be placed
on the concentration parameter, which is generally degenerate without
any additional strong lensing constraints); 2)~the velocity dispersion
$\sigma_{\mathrm{V}}$ derived from fitting SIS profiles.  Although the
details of the mass profiles are not well constrained by the weak-lensing
analysis, in particular close to the center, the total mass is robust with
respect to variations in the concentration parameter of the NFW profile.

We believe that a number of limitations are still affecting our lensing
mass estimates: 1)~we have only limited knowledge of the redshift
distribution of the galaxies used to measured the shear signal -- this
may plague the measurement of the shear signal towards the cluster center,
although the narrow redshift range of the sample ($\Delta z/z \sim 10\%$)
and the homogeneity of the weak lensing data make a direct comparison
between the different clusters possible; 2)~we could recover the shear
signal at best to the 20\% level due to a relatively small usable
number density of background galaxies (about 10~galaxies per square
arc-minute after taking into account a color selection) and a probably
not perfect PSF correction; 3)~our lensing mass determination assumes
circular symmetry which may in some cases be a poor representation of
the cluster morphology. Fortunately, thanks to the large spatial extent
of the weak-lensing detection the mass-sheet degeneracy in our mass
measurements is minimized.

As for the total masses $M_{200}$ of the clusters in our sample, we
investigated the relations between mass, X-ray and optical
observables.  Unlike most previous attempts at calibrating these
relations, we use masses measured directly from their gravitational
effects employing methods and data that are completely independent of
the X-ray measurements. In particular, the errors in the X-ray
observables and in the weak-lensing masses are uncorrelated. We
summarize the main conclusions and results of the study:
\begin{itemize}
\item The optical $M/L$ ratio presents a strong correlation with the
  cluster luminosity, with $M/L \propto L^{0.80 \pm 0.24}$. The most
  massive and luminous clusters thus have the highest $M/L$ ratio.
  This reflects a change in galaxy formation efficiency in rich
  clusters.
\item There is a strong correlation between mass and X-ray
  luminosity with $L_{\mathrm{X}} \propto M_{200}^{0.83 \pm 0.11}$.
  The logarithmic slope is significantly smaller than found in
  previous attempts to compare both quantities. A better
  understanding of the behavior of this relation is crucial in view
  of future large X-ray surveys of clusters for most of which only
  X-ray luminosities will be available.
\item The mass range of our cluster sample is too small to correctly
  fit a $M-T$ relation so we fix the logarithmic slope to $3/2$ and
  concentrate on determining the normalization to find $M_{200}/
  10^{14} M_\odot = 0.473 \, h_{70}^{-1} \,\left( T_{\mathrm{X}} / 1
  \mathrm{keV} \right) ^{3/2}$. This normalization is very close to
  the value predicted from numerical cosmological simulations of
  cluster formation and evolution as well as the observed
  normalization from X-ray measurements
  \citep{ascasibar06,springel05}. This good agreement also suggests
  that evolutionary effects are negligible between $z=0$ and $z=0.2$.
\item The scatter in the $M-T$ relation is still large and difficult
  to disentangle from uncertainties in the measured masses introduced
  by limitations in the current lensing analysis. Analysis of larger
  samples of clusters with, ideally, better weak-lensing data and
  comparison with simulated data will be required to conclusively
  address the impact of the hydrodynamical state of clusters and the
  reliability of X-ray measurements. It is, however, already clear
  from this study that our eleven clusters differ significantly in
  terms of global morphology, dynamical state, mass concentration, and
  thus possibly merging histories.
\end{itemize}

In the near future, progress on this kind of analysis may be achieved in
different ways.  Better constraints on mass profiles and particularly
on the concentration $c$ will be discussed in a forthcoming combined
analysis of strong- and weak-lensing data where the number and positions
of strongly lensed or multiply imaged galaxies unambiguously determines
the slope of the mass profile in the cluster cores (Hudelot et al., in
preparation). First results have already been obtained in Abell\,1689
\citep{limousin07} and Abell\,68 \citep{richard07}.  Analysis of larger
samples of clusters with current facilities may provide clues on cluster
physics by statistically reducing the scatter in the various scaling
relations.  However, the final limitation may arise from the intrinsic
scatter of the X-ray properties (non-equilibrium processes, cluster
mergers which tend to increase both X-ray temperature and luminosity)
or the lack of accuracy in our lensing measurements.  In conclusion,
better lensing measurements will likely be obtained, in the near
future, with very deep multi-color imaging using wide field cameras
(Subaru/SuprimeCam or CFHT/Megacam) or in the more distant future using
a space-based wide field imager such as the SNAP telescope.

\begin{acknowledgements}
  We are grateful to Sarah Bridle and Phil Marshall for many
  interactions and helpful discussions, especially regarding
  \textsc{Im2shape} and \textsc{LensEnt2}, their distribution and
  their use. The referee Thomas Reiprich helped us improve the paper
  by a lot of useful comments.  We wish to thank CALMIP (\emph{CALcul
  en MIdi-Pyr\'en\'ees}) for making data-processing resources
  available during the last quarter of 2002 for this CPU-time and RAM
  consuming analysis. We also thank the Programme National de
  Cosmologie of the CNRS for financial support. JPK acknowledges
  support from CNRS and Caltech. IRS and GPS acknowledge support from
  the Royal Society.
\end{acknowledgements}


\appendix

\section{Shear profiles of the cluster sample}
\label{app:shear-profiles}
\begin{figure*}
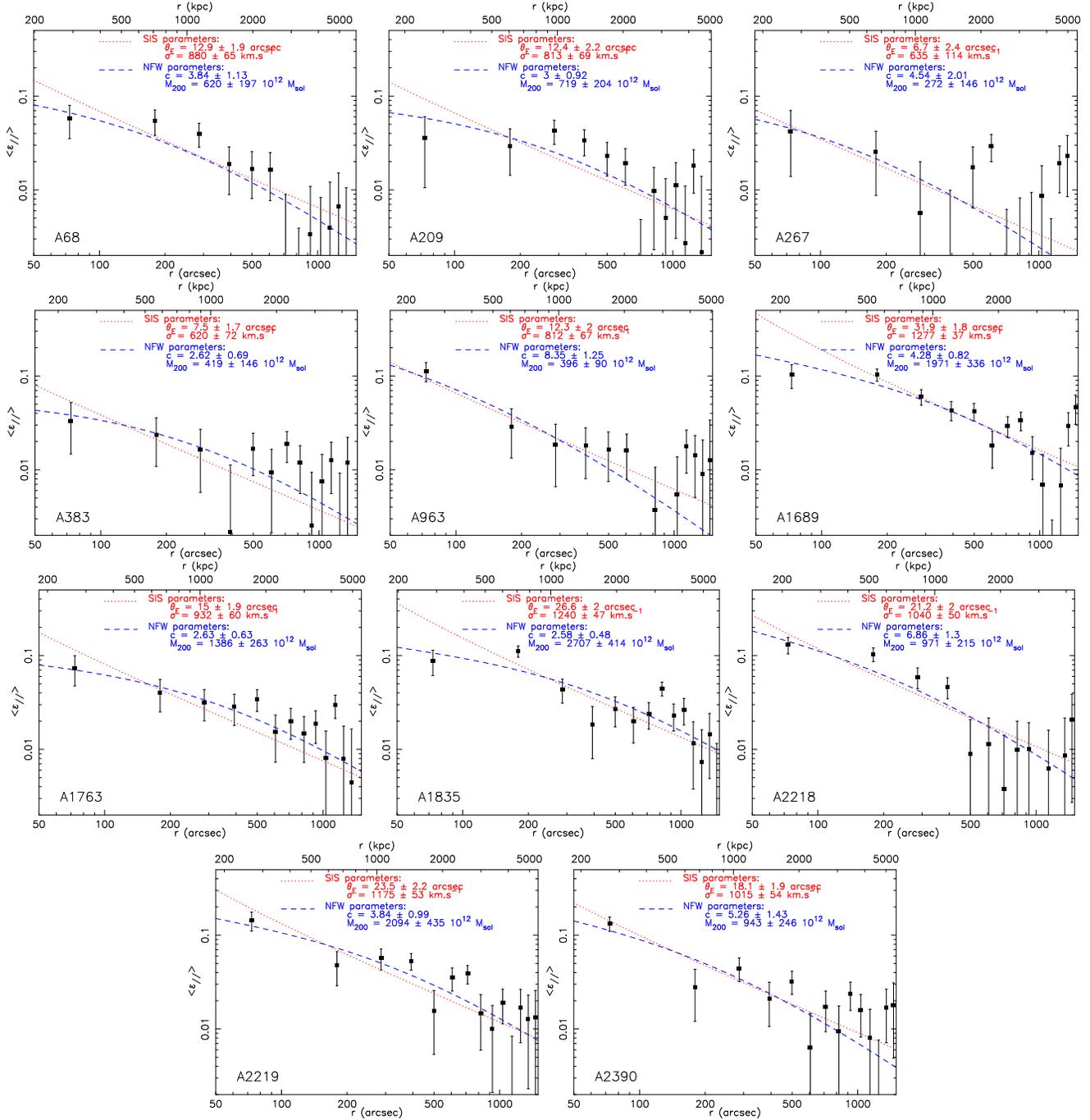

  \centering
  \includegraphics[width=0.25\textwidth,angle=-90]{Figures/A68_model2.ps}
  \includegraphics[width=0.25\textwidth,angle=-90]{Figures/A209_model2.ps}
  \includegraphics[width=0.25\textwidth,angle=-90]{Figures/A267_model2.ps}
  \includegraphics[width=0.25\textwidth,angle=-90]{Figures/A383_model2.ps}
  \includegraphics[width=0.25\textwidth,angle=-90]{Figures/A963_model2.ps}
  \includegraphics[width=0.25\textwidth,angle=-90]{Figures/A1689_model2.ps}
  \includegraphics[width=0.25\textwidth,angle=-90]{Figures/A1763_model2.ps}
  \includegraphics[width=0.25\textwidth,angle=-90]{Figures/A1835_model2.ps}
  \includegraphics[width=0.25\textwidth,angle=-90]{Figures/A2218_model2.ps}
  \includegraphics[width=0.25\textwidth,angle=-90]{Figures/A2219_model2.ps}
  \includegraphics[width=0.25\textwidth,angle=-90]{Figures/A2390_model2.ps}
  \caption{Shear profiles of the 11 clusters, using the ``red galaxy''
    catalog. The results of the fits with two different mass profiles
    (SIS and NFW) using \texttt{McAdam} are shown.}
  \label{fig:shear_prof}
\end{figure*}

The shear profiles for all clusters are shown in
Fig.~\ref{fig:shear_prof}. They are based on the ``red galaxy''
catalog and feature statistically independent data points. We also
plot the results of the \texttt{McAdam} fits obtained with the two
mass profiles discussed in the main body of this paper (SIS and NFW,
see Table~\ref{tab:mass_bestfit}).  The shear signal in the central
bin is generally similar to or weaker than the signal in the second
bin. This is likely due to residual contamination from cluster members
or foreground galaxies in the catalogs.

\section{Individual properties of the clusters}
\label{app:indiv-clusters}

\subsection{Abell\,68}
Abell\,68 contains a large cD galaxy elongated in the NW-SE direction.
About 1\arcmin\ to the North-West of the cD is a compact group of
about five bright galaxies, which gives the system a bimodal
appearance. Many blue arclets can be seen around the cluster center
and \citet{smith05} provide several redshift identifications for these
objects.  Most notable is a triple arc located just east of the center
of the cD galaxy.  Recent integral-field spectroscopic observations
revealed very strong Lyman-$\alpha$ emission corresponding to an
extended object at $z \simeq 2.63$ \citep{richard07}. This dedicated
spectroscopic survey of all the lensed galaxies in this cluster also
provides a catalog of 27~images with spectroscopic identification and
redshifts ranging from 0.4~up to 5.4, which considerably improves the
central part of the mass model. The strong-lensing model and the
weak-lensing signal appear to be in good agreement in this cluster
although they are in disagreement with the previously measured
weak lensing mass of \citet{dahle02}.

\subsection{Abell\,209}
Abell\,209 is dominated by a bright cD galaxy that is elongated in the
NW-SE direction. There are no obvious giant arc systems, although
\citet{dahle02} mention an arc candidate embedded in the envelope of
the central galaxy. The internal dynamics of the cluster has been
studied in detail by \citet{mercurio03} who find a high value of the
line-of-sight velocity dispersion ($\sigma_{\mathrm{los}} \simeq
1400\,\mathrm{km\,s^{-1}}$). The presence of a velocity gradient along
the main extension of the distribution of the cluster galaxies and
evidence for substructure and dynamical segregation suggest that we
are observing this system in a late merger phase. This interpretation
is supported by the recent detection of a radio halo
\citep{giovannini06} which implies that this cluster is a dynamically
immature merger. An independent weak-lensing analysis of Abell\,209
was performed by \citet{paulin07} using the same data retrieved from
the CFHT archives. They measured a virial mass $M_{200} =
7.7^{+4.3}_{-2.7} \times 10^{14}\,M_\odot$ inside a virial radius
$r_{200} = (1.8 \pm 0.3)\,\mathrm{Mpc}$, values that are in good
agreement with our own measurements although they were obtained with a
different weak-lensing pipeline and methodology. They are also
compatible with previous estimates \citep{dahle02}. Their results also
show a strong elongation of the cluster.

\subsection{Abell\,267}
Abell\,267 is very similar in appearance to Abell\,209. The cluster is
dominated by a giant cD galaxy with large ellipticity.  There are no
giant arcs and no obviously lensed background galaxies
\citep{smith05}.  The weak-lensing signal is the weakest of the
sample, making this one of the least massive systems studied here.

\subsection{Abell\,383}
Abell\,383 is dominated by a nearly circular cD galaxy and shows a
rich and complex system of giant arcs and arclets
\citep{smith01,smith05}. These arcs were used to reconstruct the
central mass distribution of the cluster with high accuracy. With five
multiple-image systems identified in the cluster center and three more
faint, tentative systems, the mass distribution in the core is tightly
constrained. The central slope of the mass profile is surprisingly
steep, but is consistent with the excess mass due to the strong
cooling core which feeds baryonic mass to the cD galaxy. The
weak-lensing signal, on the other hand, does not present the same
characteristics, leading us to conclude that this cluster is highly
concentrated but not widely extended.

\subsection{Abell\,963}
The center of Abell\,963 is dominated by a cD galaxy and contains two
giant arcs to the North and the South of the cD, respectively
\citep{lavery88}. \citet{ellis91} measured a redshift for the northern
arc of $z=0.771$, while the southern arc, of very blue color, has not
been identified spectroscopically. From their strong-lensing model,
\citet{smith05} argue that the southern arc is a group of singly
imaged galaxies rather than a multiple-image system. \citet{lavery88}
measured a velocity dispersion of $1350 ^{+200}_{-150}
\,\mathrm{km\,s^{-1}}$ from 36~cluster members. The weak-lensing signal
is remarkably regular and consistent with an NFW~profile even at small
radii. The spatial distribution of the cluster members is highly
circular and characteristic of a well relaxed cluster.

\subsection{Abell\,1689}
Abell\,1689 is a well known gravitational lens.  It is a very rich and
luminous cluster, dominated by a compact group of bright galaxies and
a central giant elliptical (gE) galaxy. Deep HST/ACS images reveal
a strong over-density of faint compact galaxies in the periphery
of this galaxy \citep{mieske04}, highlighting its dominance at the
center of the cluster. A dynamical study of this cluster was performed
by \citet{teague90} who found an extremely high velocity dispersion
($\sigma_{\mathrm{los}} = 1989 \mathrm{km\,s}^{-1}$) probably arising from
a complex merger.  \citet{girardi97} confirmed this initial assumption
with a detailed study of substructure in the cluster and showed that
the measured velocity dispersion probably contains a systematic peculiar
velocity component caused by an ongoing cluster merger. X-ray properties,
such as the low gas mass fraction in this cluster, confirm this
interpretation \citep{andersson04}.  Gravitational lensing features
have been studied extensively in Abell\,1689. \citet{clowe01} and
\citet{king02} present weak-shear measurements in the cluster which
give a global mass profile close to an NFW profile, compatible with the
X-ray data, but with a total mass lower than our present measurement
by nearly a factor 2.  The most spectacular results come from the deep
HST/ACS images obtained by \citet{broadhurst05a} which display a number
of arcs and arclets, making this cluster the most spectacular cluster
lens with one of the largest Einstein radii observed in clusters.
\citet{halkola06} also built a non-parametric strong-lensing mass
model and included external weak-lensing constraints to derive a total
virial mass, with a value that is 50\% higher than our measurement.
More recently, an extensive spectroscopic survey of the arcs in this
cluster has been conducted (Richard et al. 2007b, in preparation)
yielding spectroscopic redshifts for about 2/3 of the 32 multiple image
systems identified in the cluster. Using these data, \citet{limousin07}
constructed an improved mass model combining constraints from both
strong and weak lensing. They found a value of $7.6 \pm 1.6$ for the
concentration parameter $c$, similar to what is expected from numerical
simulations \citep{bullock01}, a total virial mass $M_{200} = (1.32 \pm
0.2) \times 10^{15} M_\odot$, and a virial radius $r_{200} = (2.16 \pm
0.1) Mpc$, in excellent agreement with our results.

\subsection{Abell\,1763}
Abell\,1763 has a central cD galaxy but otherwise the cluster center
is comparatively ill defined, with ``chains'' of bright galaxies
heading off in at least three directions. There are no obvious
gravitational-arc systems in the cluster center.  The weak-lensing
signal is rather weak even close to the center resulting in a low
value for the total mass. The two-dimensional mass reconstruction
clearly shows signs of bimodality in this cluster, with the main
component centered on the cD galaxy and a second component $4\arcmin$
to the west. This second mass structure was not detected by
\citet{dahle02} because it is beyond their maximum radius of
investigation.

\subsection{Abell\,1835}
Abell\,1835 is dominated by a giant elliptical galaxy slightly
elongated in the North-South direction. The galaxy distribution is
regular and the global appearance is that of a well relaxed cluster.
A\,1835 is the most X-ray luminous cluster in the BCS catalog
\citep{ebeling00} and therefore of the present sample. From the
$L_{\mathrm{X}}-T_{\mathrm{X}}$ relation it should also be the hottest
cluster; however, its measured X-ray temperature is relatively modest
(Table~\ref{tab:sample}) and drops to about 4\,keV in the cooling-core
region \citep{mcnamara06}. From a lensing point of view, A\,1835 is a well
known strong-gravitational lens, with many very thin long gravitational
arcs seen in the HST/WFPC2 image. Several of them have been identified
spectroscopically \citep{richard03,pello04,smith06}, in particular a
controversial $z=10.0$ galaxy.  A weak lensing study of this cluster was
carried out by \citet{clowe02} with the Wide Field Imager (WFI) on the
ESO/MPG 2.2\,m telescope. The isothermal fit agrees well with our work,
as does the NFW $r_{200}$ value which gives a total mass of the cluster
in very close agreement with our measurement.

\subsection{Abell\,2218}
Abell\,2218 is arguably one of the most famous cluster lenses, with an
extraordinary number of arcs and arclets in its center.  A lens model
for this cluster was presented by \citet{kneib96} and required a bimodal
central mass distribution, with one mass component centered on the cD
galaxy and a second one centered on a bright galaxy about 1.5\arcmin\
to the south-east of the cD. \citet{girardi97} analyzed the distribution
of 50 galaxy redshifts in A\,2218 and found evidence for two groups of
galaxies superimposed along the line of sight, which they identify with
the mass clumps modeled by \citet{kneib96}. Deep HST/ACS images revealed
the nature of several strongly lensed galaxies and in particular a $z
\sim 7$ galaxy candidate, recently confirmed by Spitzer observations
\citep{kneib04,egami05}. The weak-lensing signal detected in this
cluster is one of the strongest in our sample, and the two-dimensional
mass distribution is perfectly matched to the galaxy distribution, with
an ellipticity and orientation which reflects the central bimodality of
the mass distribution.

\subsection{Abell\,2219}
The optical morphology of Abell\,2219 is remarkably similar to that of
Abell\,2218, with a dominant cD galaxy and a second bright elliptical
galaxy at $\sim 1\arcmin$ to the South-West.  A number of gravitational
arcs can be seen, most notably a straight arc between the two brightest
cluster galaxies and a very thin and elongated arc to the North-West. The
presence of two mass clumps as well as the elongated X-ray distribution
indicate a non-relaxed cluster in the process of merging of several
substructures. A gravitational depletion signal at near-infrared
wavelengths was detected by \citet{gray00}.  The mass deduced from
this effect was fitted with a singular isothermal mass distribution
with $\sigma \sim 800\,\mathrm{km\,s^{-1}}$ and is consistent with our
measurement.  However, our measurement is slightly lower than the one
obtained by \citet{dahle02}. The two-dimensional mass reconstruction
reflects the general elongation of the cluster, although this is less
significant than in Abell\,2218, for example.

\subsection{Abell\,2390}
Abell\,2390 is a cD-dominated cluster with several arcs in its center.
A chain of fairly bright galaxies extends to the North-West of the cD
galaxy. The arcs on this side of the cD are straight \citep{pello91}
and confirm the extension of the underlying mass distribution in
this direction. About $3\arcmin$ to the East of the cluster center
lies an extended group of galaxies. This cluster has a surprisingly
high velocity dispersion \citep{leborgne91,yee96,borgani99}. X-ray
observations reveal a strongly elongated gas distribution along the
NW-SE axis \citep{pierre96,allen01a}, consistent with the optical galaxy
distribution. Our own two-dimensional mass reconstruction has a similar
elongation in the same direction supporting the notion that A\,2390 is a
non-relaxed cluster, accreting one or two mass clumps along the same axis.

\end{document}